\documentclass{aa}
\usepackage{graphicx}
\usepackage{txfonts}
\usepackage{lscape}
\usepackage{natbib}
\bibpunct{(}{)}{;}{a}{}{,}

\def\XMM{{\em XMM$-$Newton}}

\def\ergscm2{\rm erg\,cm^{-2}\,s^{-1}}

\begin{document}

\title{Characterization of new hard X-ray Cataclysmic 
Variables\thanks{Based on observations obtained with 
\XMM\ and INTEGRAL, ESA science missions with instruments and contributions 
directly funded by  ESA Member States; and Swift, a NASA science 
mission with Italian participation.}}

\author{F.~Bernardini\inst{1}
\and
D.~de Martino\inst{1}
\and
M.~Falanga\inst{2}
\and
K.~Mukai\inst{3}
\and
G.~Matt\inst{4}
\and
J.-M.~Bonnet-Bidaud\inst{5}
\and
N.~Masetti\inst{6}
\and
M.~Mouchet\inst{7}
}


\institute{
INAF - Osservatorio Astronomico di Capodimonte, salita Moiariello 16, I-80131 Napoli, Italy \\
\email{federico.bernardini@oa-roma.inaf.it}\\
\email{demartino@oacn.inaf.it}
\and
International Space Science Institute (ISSI), Hallerstrasse 6, 
CH-3012 Bern, Switzerland \\
\email{mfalanga@issibern.ch}
\and
 CRESST and X-Ray Astrophysics Laboratory, NASA Goddard Space Flight 
 Center, Greenbelt, MD 20771,
 USA and Department of Physics, University of Maryland, Baltimore County, 
 1000 Hilltop Circle, Baltimore, MD 21250, USA \\
\email{koji.mukai@nasa.gov}
\and
Dipartimento di Fisica, Universit\'a Roma III, Via della Vasca Navale 84,
I-00146, Roma, Italy \\
\email{matt@fis.uniroma3.it}
\and
CEA Saclay,  DSM/Irfu/Service d'Astrophysique, F-91191 
Gif-sur-Yvette, France \\
\email{bonnetbidaud@cea.fr}
\and
INAF Istituto di Astrofisica Spaziale e Fisica Cosmica di Bologna, Via Gobetti 101, I-40129, Bologna, Italy \\
\email{nicola.masetti@iasfbo.inaf.it}
\and
Laboratoire APC, Universit\'{e} Denis Diderot, 10 rue Alice Domon et L\'{e}onie Duquet, F-75005
Paris, France and LUTH, Observatoire de Paris, Section de Meudon, 5 place Jules Janssen, F-92195
Meudon, France \\
\email{martine.mouchet@obspm.fr}
}

\date{Accepted for publication in A\&A in April 2012}

\abstract{}
{We aim at characterizing a sample of nine new hard X-ray selected
Cataclysmic Variable (CVs), to unambiguously identify them as magnetic systems of the Intermediate Polar (IP) type.}
{We performed detailed timing and spectral analysis by using X-ray, and simultaneous UV
and optical data collected by \XMM, complemented with hard X-ray data provided by
INTEGRAL and Swift. The pulse arrival time were used to estimate
the orbital periods. The broad band X-ray spectra were fitted
using composite models consisting of different absorbing columns and emission components.}
{Strong X-ray pulses at the White Dwarf (WD) spin period are detected and found to
decrease with energy. Most sources are spin-dominated systems in the
X-rays, though four are beat dominated at optical wavelengths.
We estimated the orbital period in all system (except for
IGR~J16500-3307), providing the first estimate for
IGR\,J08390-4833, IGR\,J18308-1232, and IGR\,J18173-2509.
All X-ray spectra are multi-temperature. V2069 Cyg and
RX~J0636+3535 posses a soft X-ray optically thick component
at kT$\sim$80 eV. An intense K$_{\alpha}$ Fe line at 6.4 keV is detected in all sources.
An absorption edge at 0.76 keV from OVII is detected in IGR~J08390-4833.
The WD masses and lower limits to the accretion rates are also estimated.}
{We found all sources to be IPs. IGR~J08390-4833, V2069 Cyg,
and IGR~J16500-3307 are pure disc accretors, while
IGR~J18308-1232, IGR~J1509-6649, IGR~J17195-4100, and RX~J0636+3535
display a disc-overflow accretion mode. All sources show a temperature gradient in the
post-shock regions and a highly absorbed emission from material located in the
pre-shock flow which is also responsible for the X-ray pulsations.
Reflection at the WD surface is likely the origin of the
fluorescent iron line. There is an increasing evidence for the presence of a warm absorber in
IPs, a feature that needs future exploration. The addition of two systems to the
subgroup of soft X-ray IPs confirms a relatively large ($\sim 30\%$) incidence.}


\titlerunning{New IPs}

\maketitle

\section{Introduction}

Magnetic Cataclysmic Variables (CVs) constitute a 
subgroup of the CV class, harboring  accreting white dwarfs (WD) with  magnetic field 
strengths $\rm B\ga 10^5\,G$. These systems are further subdivided in two groups, 
depending on the WD magnetic field intensity and degree of synchronism ($\rm 
P_{rot=\omega}/P_{orb=\Omega}$). Those called Polars are synchronous mCVs. They show signatures 
of strong magnetic fields ($\rm B\sim10-230$ MG) through 
the presence of a conspicuous polarization at optical and near-IR wavelengths. The 
so-called Intermediate Polars (IPs) possess instead asynchronously rotating WDs ($\rm
P_{\omega}<<  P_{\Omega}$) and except for a few cases 
\citep{piirolaetal93,potter97,katajainen07,butters09,potter2012}, they
do not show polarization. These systems are believed to possess 
weakly magnetized WDs ($\rm 
B<$10 MG). The Polars populates the orbital period distribution at short periods 
mainly below the 2--3 h orbital period "gap" \citep{warner,wheatley95}. On the 
other hand, IPs are 
generally found above the gap, with only a handful of systems below it.
 
The material lost from the Roche-lobe overflowing late-type 
secondary star (typically a Main Sequence or a sub-Giant) may flow toward 
the WD in different modes, depending on the magnetic field intensity: 
directly (stream), through a truncated disc or ring. 
The high field Polars are direct accretors, while IPs may accrete via a stream or a disc/ring 
depending on their magnetic moment and spin-to-orbital period ratio 
\citep{norton04,norton08}. A combination of these, called 
disc-overflow, may also occur \citep{hellier95,norton97} and is linked to 
the mass accretion rate.
Close to the WD, the magnetically confined flow of matter is 
accreted through a column in stream-fed systems and through an arc-shaped curtain 
\citep{rosen88} in disc/ring-fed systems. Approaching the WD surface, the velocities 
are supersonic and a stand-off 
shock is formed. In the post-shock region (PSR) the gas at a temperature of
kT$\sim10-60$ keV slows and cools via thermal 
bremsstrahlung and cyclotron radiations 
\citep{aizu73,wu94,cropper99}. The efficiency of the two 
cooling mechanisms depends on the WD magnetic  field strength 
\citep{woelk_beuermann96,fischer_beuermann01}. The post-shock emission 
is partially intercepted and thermalized by the WD surface, giving rise to an 
optically thick 
emission that emerges in the soft X-rays and EUV/UV regimes. An intense soft ($\rm kT\sim$ 
20--50 eV) component is a  characteristic signature of the X-ray spectra of the 
Polars, mostly balancing the cyclotron flux \citep{beuermann04}. 
On the other hand IPs  have typically stronger hard X-ray fluxes 
than the Polars and 
only recently a soft highly  absorbed optically thick component  has been 
detected \citep{demartino04,evans_hellier07,anzolin08,anzolin09}.

Though mCVs represent a relatively small ($\sim 20\%$) fraction of CVs, this number is 
rapidly increasing thanks to the recent hard X-ray surveys conducted by INTEGRAL and 
Swift above 20\,keV. So far 64 CVs are identified in the latest catalogue releases  
\citep{bird10,cusumano10}, 43 of them are mCVs. 
 The IPs represent $\sim 80\%$ of this hard X-ray sample (including both 
already known  \citep{barlow06} and recently discovered members 
\citep{bonnetbidaud07,demartino08,anzolin09, Butters08,Bonnet-Bidaud09,Pretorius09,scaringi11},
though one misidentification was found \citep{demartino10}. 
Hence, while in the pre-INTEGRAL and Swift times IPs amounted 
to only $\sim$20 members, this number has more than doubled nowadays. 

Hard X-ray mCVs have the potential to be important contributors to the X-ray source 
population at low luminosities ($\rm \sim 10^{30}-10^{33}\,erg\,s^{-1}$). They were proposed 
to be the major constituent of galactic ridge \citep{sazonov06,revnivtsev08,revnivtsev09} and galactic bulge
\citep{revnivtsev11,hong2012} X-ray emission. Therefore, they are also believed to have an important role in the 
X-ray luminosity function of other galaxies.  
 
In the framework of a program conducted with \XMM, with 
the main goal to identify new magnetic type CVs, we present here 
the analysis of simultaneous X-ray, UV and optical data of a sample of nine CVs 
that were detected as hard X-ray sources by INTEGRAL, RXTE or SWIFT:
IGR~J08390-4833, IGR~J18308-1232, IGR~J18173-2509, IGR~J17195-4100,
V2069 Cyg (also known as RX~J21237+4218), V647 Aur (also known as RX~J0636+3535), IGR~J15094-6649,
V515 And (also known as XSS~J0056+4548), and IGR~J16500-3307 (hereafter IGR~J0839, IGR~J1830, IGR~J1817, 
IGR~J1719, V2069 Cyg, RX~J0636, IGR~J1509, XSS~J0056, and IGR~J1650 respectively. 
We complement the X-ray analysis with a high-energy coverage from INTEGRAL/IBIS and
Swift/BAT publicly available data.
Five of these sources (IGR\,J0839, IGR\,J1817, XSS\,J0056, 
IGR\,J1509 and IGR\,J1719) have also  $Chandra$, RXTE, or Swift/XRT 
coverage.  All, except IGR\,J0839, 
IGR\,J1830 and IGR\,J1817, were also observed in optical  photometry 
and spectroscopy from ground-based telescopes, from which orbital periods were
determined and optical pulses  detected, thus allowing a comparison with our 
results.

In Sect. \ref{subs:timing} we present the X-ray, UV and optical timing analysis.
In Sect. \ref{sub:spec} we present the analysis of the 
X-ray broad-band spectra of each source. For the three brightest sources we 
also present the high resolution spectra provided by the \XMM\ RGS instrument. 
In Sect. \ref{sec:discuss} we discuss the origin of rotational pulses, 
the accretion mode and the emission properties of our sample.

\section{Observations and data analysis}
\label{sec:obs}

\subsection{\textit{XMM-Newton} observations}
 
The \XMM\ Observatory includes three
$\sim1500$~cm$^2$ X$-$ray telescopes with an EPIC instrument in each
focus, a Reflecting Grating Spectrometer, RGS \citep{denherder01}
and an Optical Monitor, OM \citep{mason01}. Two of the EPIC imaging
spectrometers use MOS CCDs \citep{turner01} and one uses a PN CCD
\citep{struder01}. \XMM\ collected data of all nine CVs. 
The main observation parameters, together with that of 
all other \textbf{observatory} instruments are reported 
in table \ref{tab:observ}.

Data were processed with SAS version
10.0.0, using the updated calibration files (CCF) available in January 2011. 
All observations were performed with the EPIC-PN (PN hereafter) camera set
in {\tt prime full window} imaging mode (time
resolution=0.0734 s) with a thin filter applied.
Standard data screening criteria were applied in the extraction
of scientific products. We accumulated a one$-$dimensional image and
fitted the 1D photon distribution with a Gaussian. Then, we extracted
the source photons from a circular region of radius 40\arcsec\
centered at the Gaussian centroid. The
background was obtained (within the same CCD where the source lies) from a circular region of 
the same size. Single and double pixel events
with a zero quality flag were selected for the PN data. 
For the spectral analysis we cleaned all observations from solar
flares by collecting CCD light curves above 10\,keV and applying an
intensity threshold. Spectra were then produced only for those parts of observations under the threshold limit. 
On the other hand, for the timing analysis we used the {\tt epiclccor} task, producing 
a background subtracted light curve in the range 0.3--15 keV (with a bin time of 11 s), 
consequently removing contamination from solar flares. 
The source event arrival time of each observation, in the 0.3$-$15\,keV
energy range, were converted into barycentric dynamical times (BDT) by
means of the SAS tool {\tt barycen}.

The PN spectra were rebinned before fitting, to have at least 30 counts per bin. 
We  report the analyses obtained with the PN data only (consistence with the results of
EPIC-MOS cameras was always verified). Phase--resolved spectra were also extracted 
at the pulse maximum and minimum. All spectra were analyzed using the version of $XSPEC$ 
(12.5.1n).

We also inspected the RGS spectra. We used the 1st order spectra and 
responses
produced by the pipeline, which runs an automatic version of the RGS
meta-task,  {\tt rgsproc}.  While, in principle, the quality of data products
can be improved somewhat by manually repeating the reduction, we found
no obvious problems with the pipeline products.  Also considering the
limited statistical quality of the RGS data for our sample, we believe
our choice to use the standard pipeline products is well justified.
The RGS spectra of IGR\,J1719, IGR\,J1509, and XSS\,J0056
were of high enough statistical quality to merit further analysis;
these are discussed below in sect. \ref{subsub:rgs}.  The data 
for  the other targets were of too low S/N to add useful 
information.
We grouped the RGS spectra of the three sources  such that each new bin 
had at least 25 counts.

Simultaneous coverage to the X-rays was ensured by the OM operated in 
fast window mode using sequentially the B (3815-4910\,\AA) or the U 
(3000--3800\,\AA) filter and UVM2 (2000--2800\,\AA)  or 
UVW1 (2450--3200\,\AA) filter. Consequently, each target was observed for 
half of the time of the EPIC exposure in the B or U filter and the other half in a UV 
filter. Due to the OM fast window mode acquisition, the OM photometric data  in each filter 
consist of sequential (four or five) segments of length $\sim$ 1300-2800\,s each (according to the 
total EPIC exposures). Average net count rates and instrumental magnitudes  are reported for 
those sources  with usable data in table\,\ref{tab:observ}, where we also report 
whether no variability is detected.
Background subtracted OM light curves were obtained using the standard 
SAS processing pipeline with a binning time of 10\,s. Barycentric corrections were also 
applied to the OM light curves.

\begin{table*}
\caption{Summary of main observations parameters for all instruments.}
{\tiny
\begin{center}
\begin{tabular}{cccccccc}
\hline\hline
 & & & & & & & \\
 
Source & Telescope      & OBSID & Instrument$^*$ & Date            & UT$_{\rm start}$ & T$_{expo}$  &Net count rate \, (mag)$^{***}$\\
       &                &       &                & yyyy-mmm-dd      & hh:mm            & (ks)$^{**}$ &    cts/s                                  \\
 & & & & & & & \\

\hline

 & & & & & & & \\
IGR~J08390-4833 & \emph{XMM-Newton}&  0651540101 & EPIC-pn & 2010 Dec 20 & 12:34 & 33.9 (27.3)&  $0.776\pm0.006$ \\
         &  & & OM-B 	&   & 12:20	& 15.0	& $3.6 \pm 0.9$ (17.8) \\
	 &  & & OM-UVM2 &            & 16:57  & 15.2 	& $0.25 \pm 0.05$ (17.3)\\
           &  \emph{INTEGRAL} & & IBIS/ISGRI &  &	& 2000 & $0.26\pm0.04$ \\ 
 & & & & & & & \\
IGR~J18308-1232 & \emph{XMM-Newton}& 0601270501 &  EPIC-pn & 2010 Mar 11 & 15:26 & 27.6 (15.7)& $0.793\pm0.007$ \\
           &  \emph{INTEGRAL} &  & IBIS/ISGRI &  & & 2200 &  $0.36\pm0.05$ \\
  & & & & & & & \\
IGR~J16500-3307 & \emph{XMM-Newton}& 0601270401  &EPIC-pn & 2010 Feb 2 & 11:04 & 30.0 (18.0)& $1.33\pm0.01$\\
              &  & & OM-B$^a$   & & 10:50 & 6.2 & $10\pm2$ (16.7)\\
& & & OM-UVW1$^a$ & & 13:00  & 8.0 & $2.2\pm0.5$ (16.3) \\
       &  \emph{INTEGRAL} & & IBIS/ISGRI &  & & 2200  & $0.31\pm0.02$ \\
           & &         &      &      &        &          & \\
IGR~J18173-2509 & \emph{XMM-Newton}& 0601270301 &EPIC-pn & 2009 Sep 07 & 02:03 & 35.0 (30.0)& $0.941\pm0.006$\\
         &  & & OM-B 	&  & 01:49 & 15.7	& $3.0 \pm 1.0$  (18.1)\\
           &  \emph{INTEGRAL} & &IBIS/ISGRI &  & & 4500  & $0.88\pm0.04$ \\
         &  & &         &            &        &          & \\
IGR~J17195-4100 & \emph{XMM-Newton}& 0601270201 & EPIC-pn & 2009 Sep 03 & 06:58 & 24.0$^a$ (27.0) & $6.70\pm0.02$ \\
      &  & & OM-B     & & 06:44 & 11.4 & $36\pm5$ (15.4)\\ 
      &  & & OM-UVM2$^b$    & & 06:44 & 11.4 & $2.3\pm0.6$ (14.9)\\ 
      &  & & RGS 1\&2     &  &06:35 &     33.9      & \\
      &  \emph{INTEGRAL} & & IBIS/ISGRI &  & & 2600 & $0.93\pm0.05$ \\
      &  & &       &     &        &          & \\

V2069 Cyg &  \emph{XMM-Newton}& 0601270101 & EPIC-pn & 2009 Apr 30 & 11:09 & 26.4 (12.4)& 1.05$\pm$0.01 \\
  &   & & OM-B     & & 10:55 & 10.6 & 10$\pm$1.5 (16.8)\\
   &   & & OM-UVM2$^a$     & & 14:18 & 10.6  & $0.17\pm0.04$ (17.6) \\
        &  \emph{INTEGRAL} & & IBIS/ISGRI &  & & 1900  & $0.14\pm0.05$ \\
           & &         &            &        &    &      & \\
RX~J0636+3535 &   \emph{XMM-Newton} & 0551430601 & EPIC-pn & 2009 Mar 18 & 17:01 & 29.2 (12.1) & $1.37\pm0.01$ \\
   &     & & OM-B$^a$     & & 16:47 & 6.7 & $13\pm2$ (16.4)\\
    &  & & OM-UVM2$^a$     & & 22:41 & 6.7  & $1.5\pm0.2$ (15.4) \\
       &  \emph{Swift} & & BAT &  & & $\sim7500$ & $0.00028\pm0.00003$ \\
           & &         &            &        &   &       & \\
IGR~J15094-6649 &  \emph{XMM-Newton}& 0551430301 &  EPIC-pn & 2009 Feb 02 & 13:39 & 30.0 & 2.56$\pm$0.01\\
   &  & & OM-U     & & 13:25 & 14.2 & $36\pm4$ (14.3)\\
   &   & & OM-UVM2$^a$    & & 17:49 & 12.4  & $2.8\pm0.8$ (14.7) \\
      &  & & RGS 1\&2     &  & 13:16 &  31.9        & \\
       &  \emph{INTEGRAL} & & IBIS/ISGRI &  & & 1400 & $0.61\pm0.06$ \\
           & &         &            &        & &         & \\
XSS~J0056+4548 & \emph{XMM-Newton}& 0501230301 & EPIC-pn & 2007 Dec 31 & 04:56 & 15.0 (7.4)& $3.06\pm0.02$\\
 &  & & OM-B     & & 04:42 & 6.7 & $69\pm5$ (14.6) \\
 & & & OM-UVM2 & & 07:01 & 6.7 & $10\pm1$ (13.2) \\
      &  & & RGS 1\&2     &  & 04:33 &   16.9       & \\
       &  \emph{INTEGRAL} & & IBIS/ISGRI &  & & 240 & $0.35\pm0.06$ \\
           & &         &      &      &        &          & \\
\hline
\hline
\end{tabular}
\label{tab:observ}
\end{center}
}
$^{*}$ \emph{Optical Monitor} band reported when data usable\\
$^{**}$ Net exposure times. In parenthesis is reported the solar flare removed exposure \\ 
$^{***}$ Instrumental magnitude reported in parenthesis.\\
$^a$ No variability detected.\\
$^b$ Time window filter applied: only first 2/3 of observation are used.\\
\end{table*}

\subsection{The INTEGRAL observations}
 
 The INTEGRAL IBIS/ISGRI instrument \citep{ubertini03,lebrun03} observed all sources except 
RX\,J0636.
 Hard (20--100 keV) X-ray data were extracted from all pointings within 12\degr\ from the 
source positions. 

To study the persistent X-ray emission, the time-averaged 
ISGRI spectra were obtained from mosaic images in five energy bands, 
logarithmically spaced  between 20 and 100 keV.
Data were reduced with the standard OSA software version 7.0 and then 
analyzed with  the algorithms 
described by \cite{goldwurm_etal03}.

\subsection{The Swift observations}

 The Swift Burst Alert Telescope, BAT \citep{Barthelmy}, is a wide-field ($\sim$1 steradian) 
coded aperture mask
instrument sensitive in the 14--195 keV range.  Thanks to the large field of view, BAT
has built up a sensitive all-sky map of the hard X-ray sky.  The BAT team provides
the average 8-channel spectrum of each source detected over the first 58 month of
the mission. 

RX J0636 only recently appeared in the Swift/BAT 58 month survey catalogue.
To obtain an high-energy coverage of this source we downloaded the BAT spectrum 
along with the response matrix (http://swift.gsfc.nasa.gov/docs/swift/results/bs58mon/).

\section{Results}
\label{results}

\subsection{Timing analysis}
\label{subs:timing}

All sources show periodic-like variability in their EPIC light 
curves and six of them also in their OM light curves. We then analyzed these light curves 
to identify periodic signals and to characterize them at different energies.

\subsubsection{X-ray variability}

\begin{figure*}
\begin{center}
\begin{tabular}{cc}
\includegraphics[width=3.0in, height=3.0in, angle=-90]{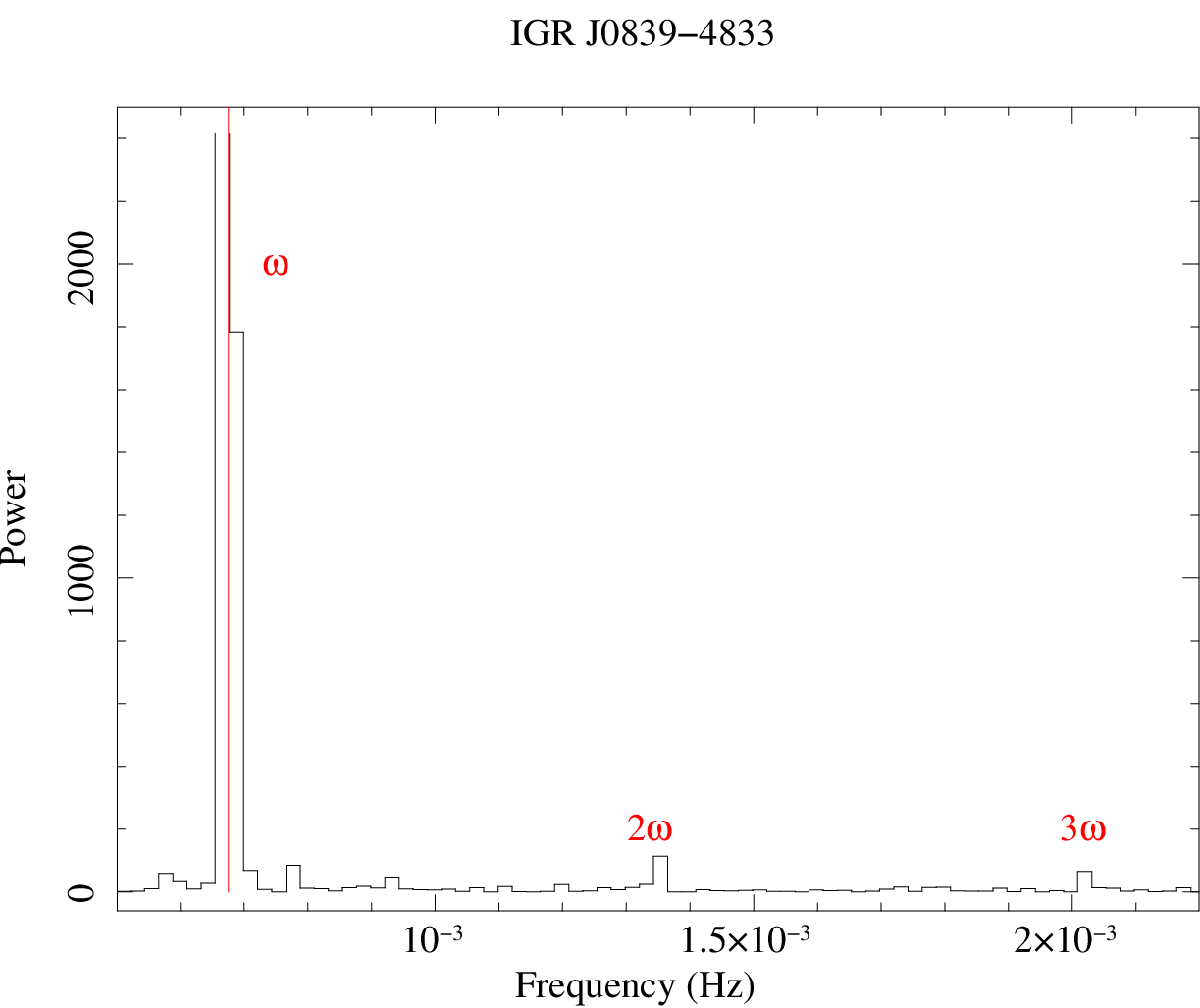} & \includegraphics[width=3.0in, height=3.0in, angle=-90]{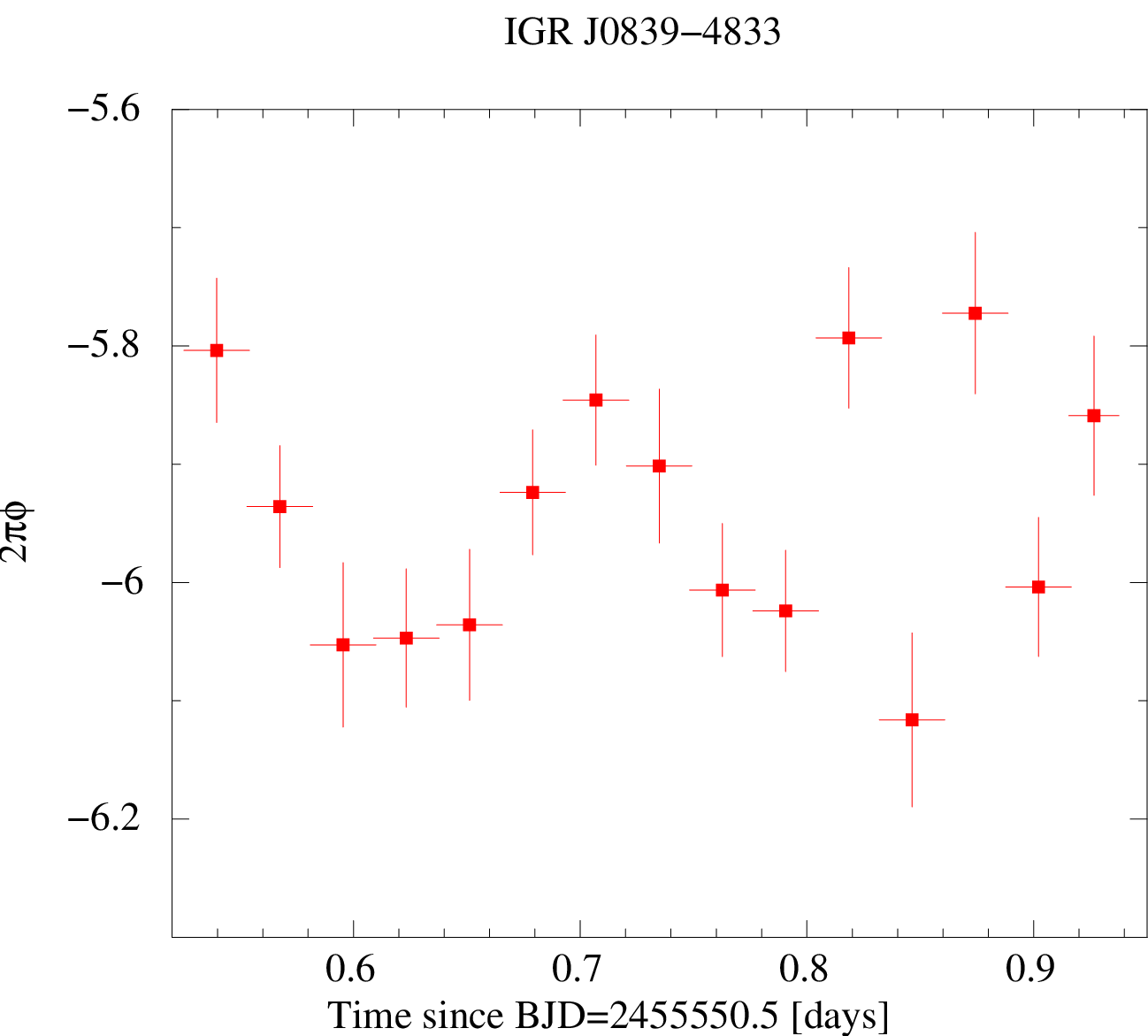}\\ 
\includegraphics[width=3.0in, height=3.0in, angle=-90]{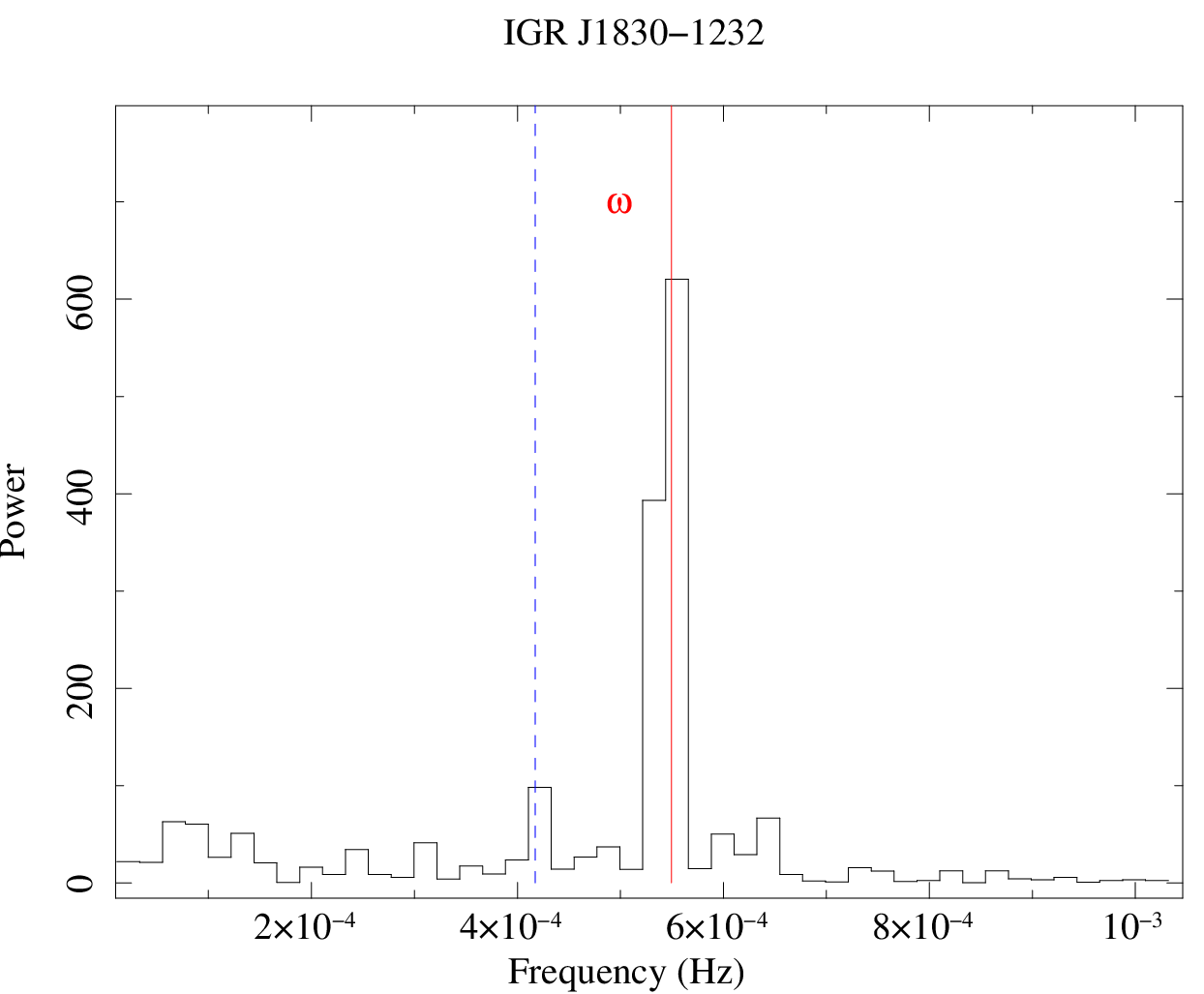} & \includegraphics[width=3.0in, height=3.0in, angle=-90]{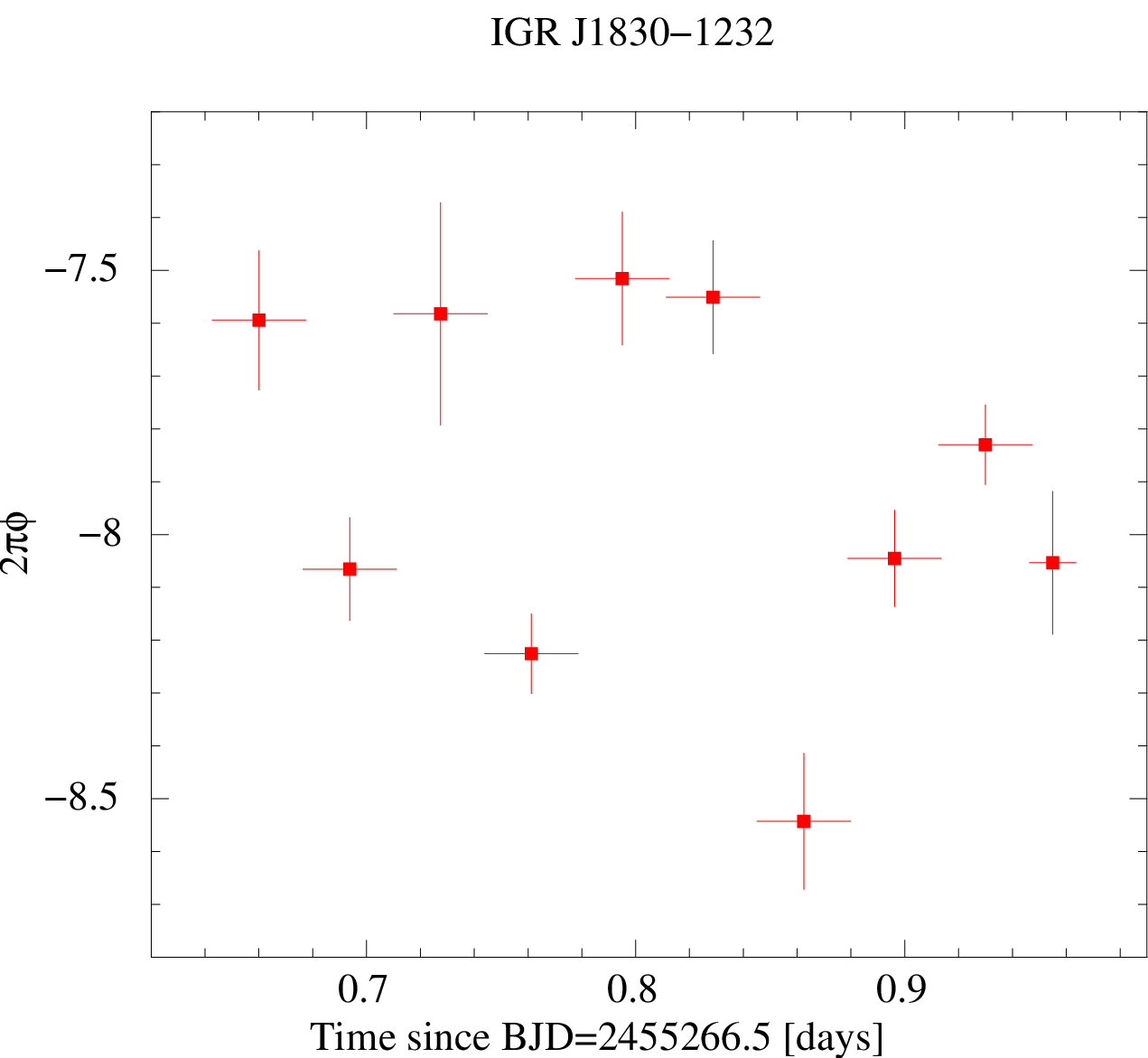}\\ 
\includegraphics[width=3.0in, height=3.0in, angle=-90]{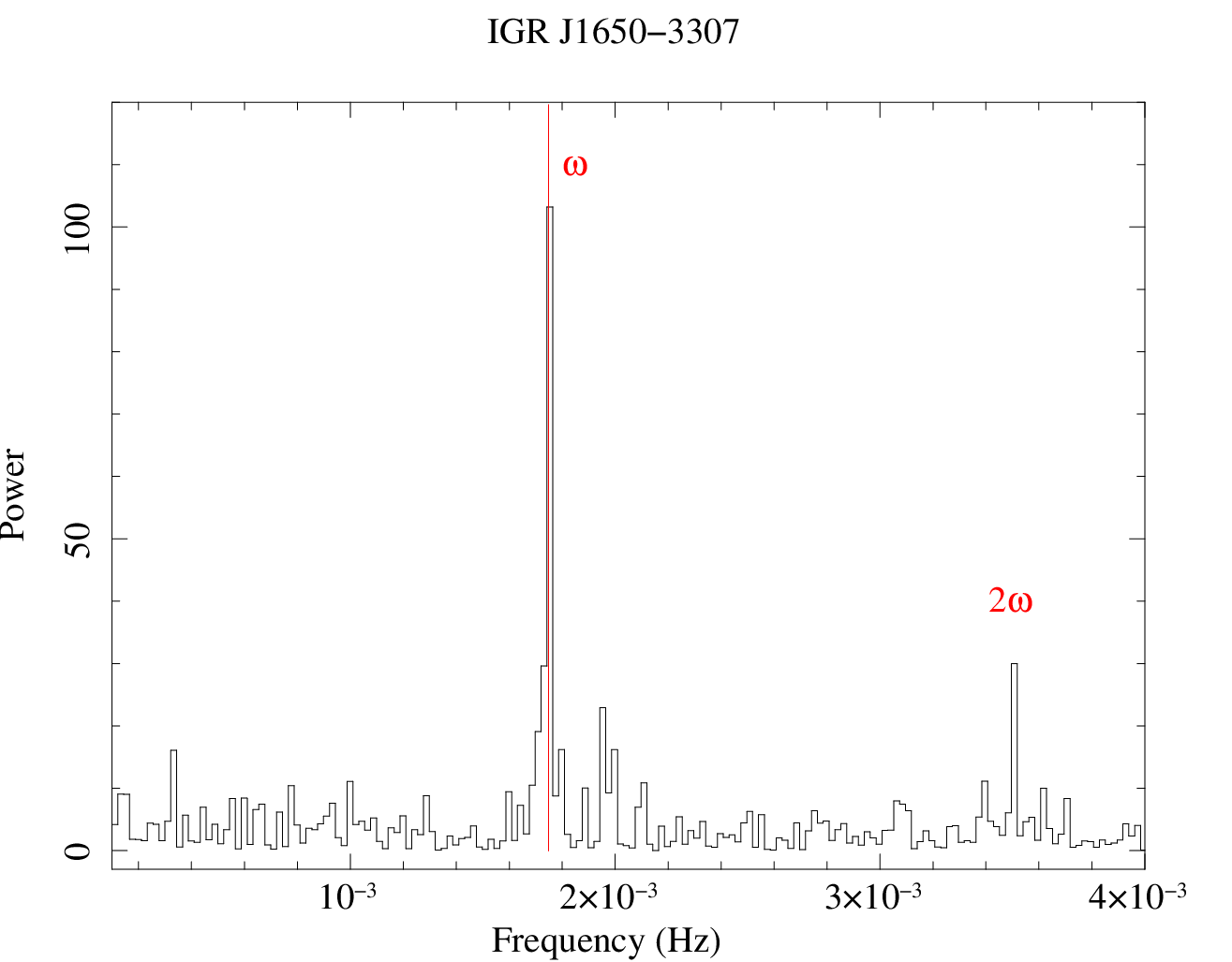} & \includegraphics[width=3.0in, height=3.0in, angle=-90]{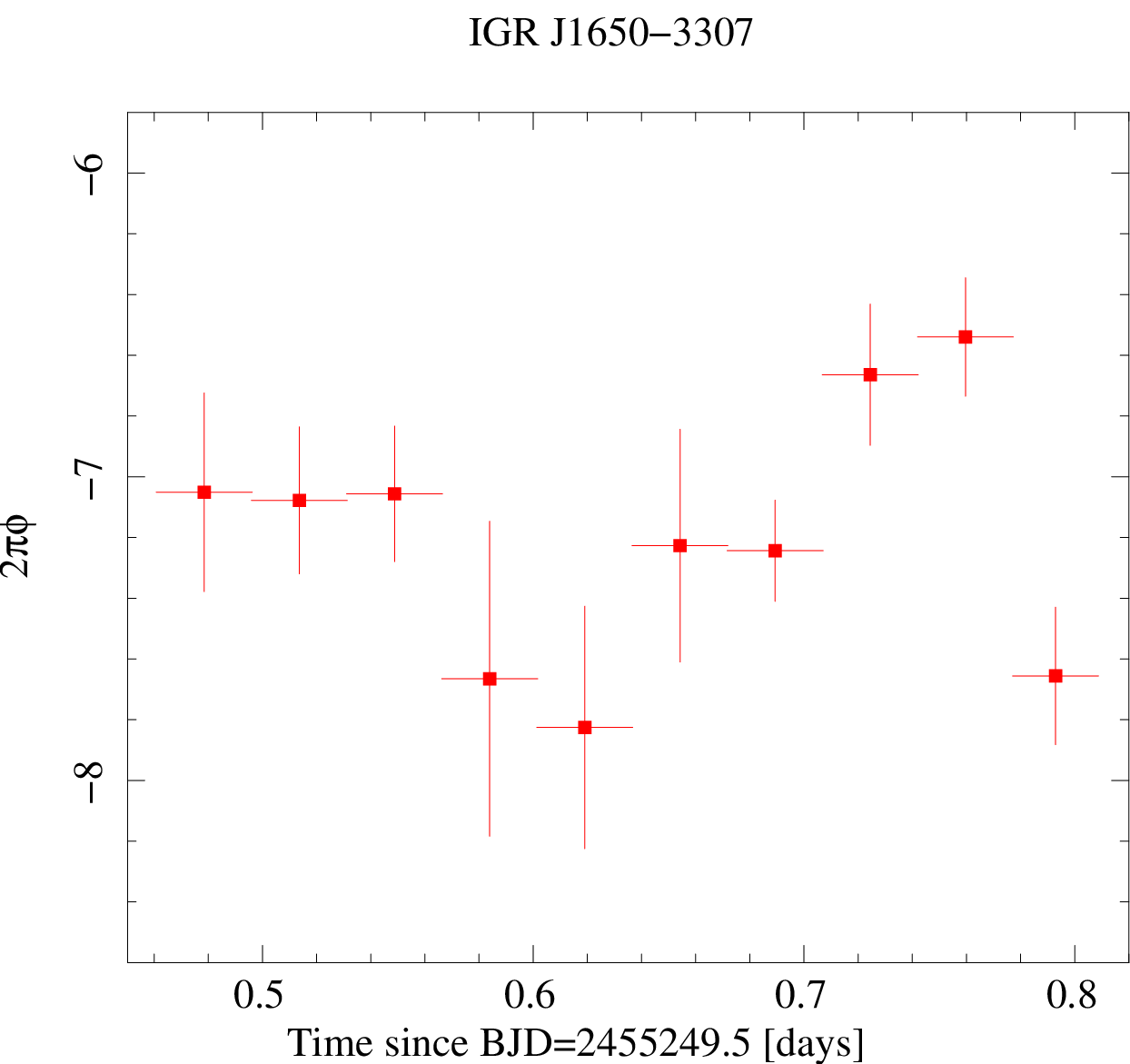} \\
\end{tabular}
\caption{\textit{Left Panel}: Power spectrum. Fundamental and  harmonics 
are reported in red, while dotted blue lines represent sidebands. 
\textit{Right panel}: Evolution of the phase of the main signal, $\omega$, vs time.}
\label{fig:ps_porb1}
\end{center}
\end{figure*}

\begin{figure*}
\begin{center}
\begin{tabular}{cc}
\includegraphics[width=3.0in, height=3.0in, angle=-90]{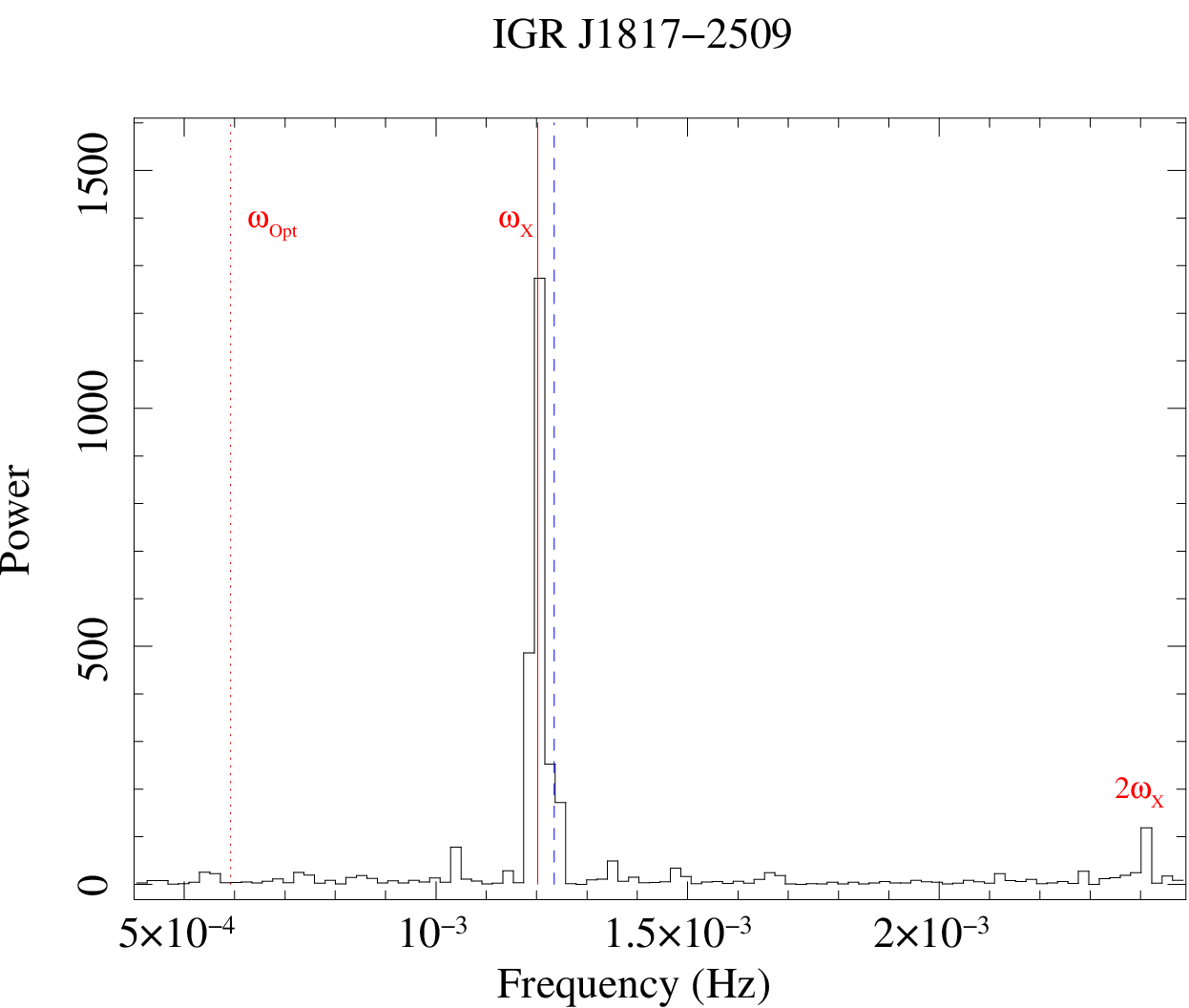} & \includegraphics[width=3.0in, height=3.0in, angle=-90]{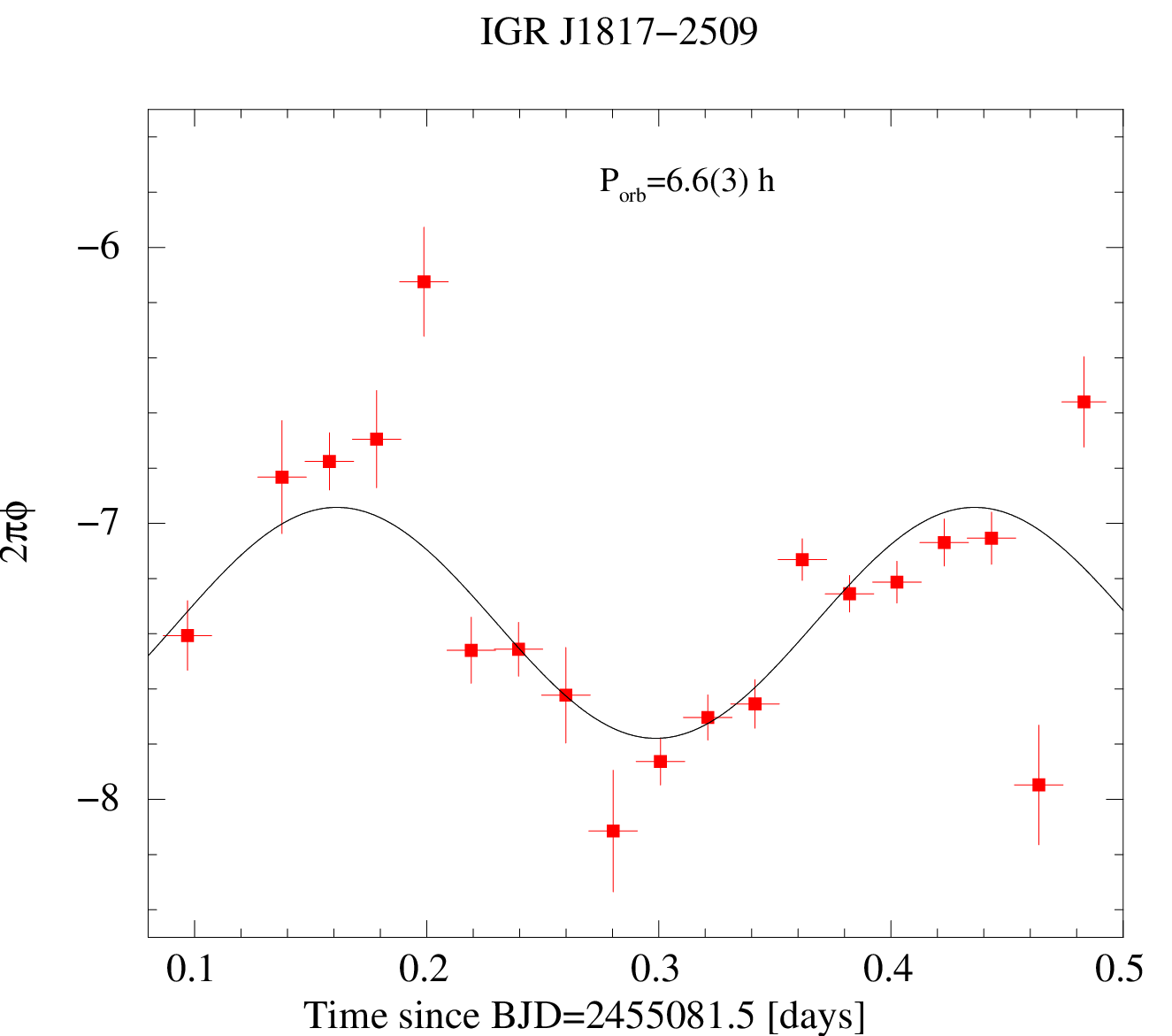}\\
\includegraphics[width=3.0in, height=3.0in, angle=-90]{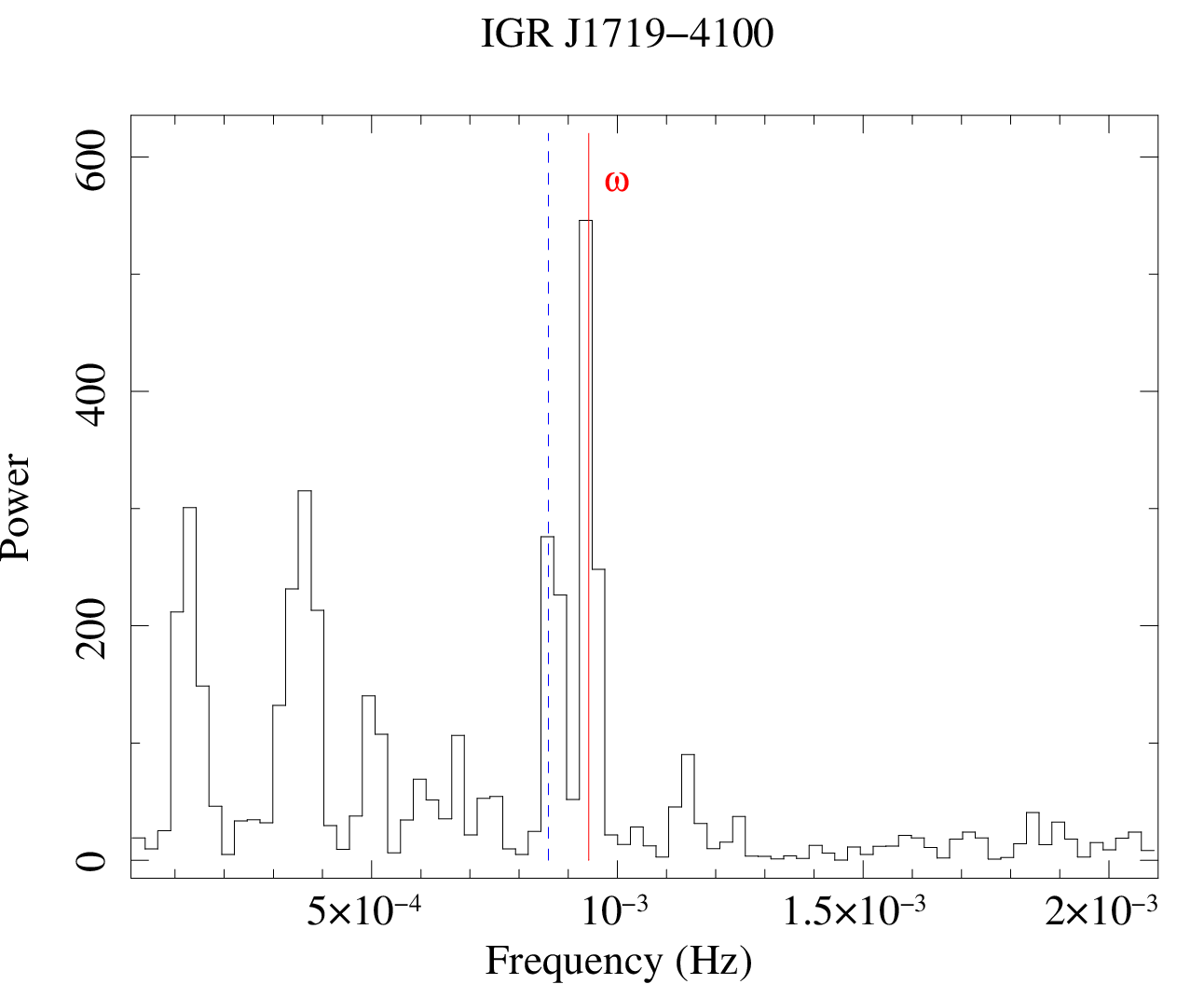} & \includegraphics[width=3.0in, height=3.0in, angle=-90]{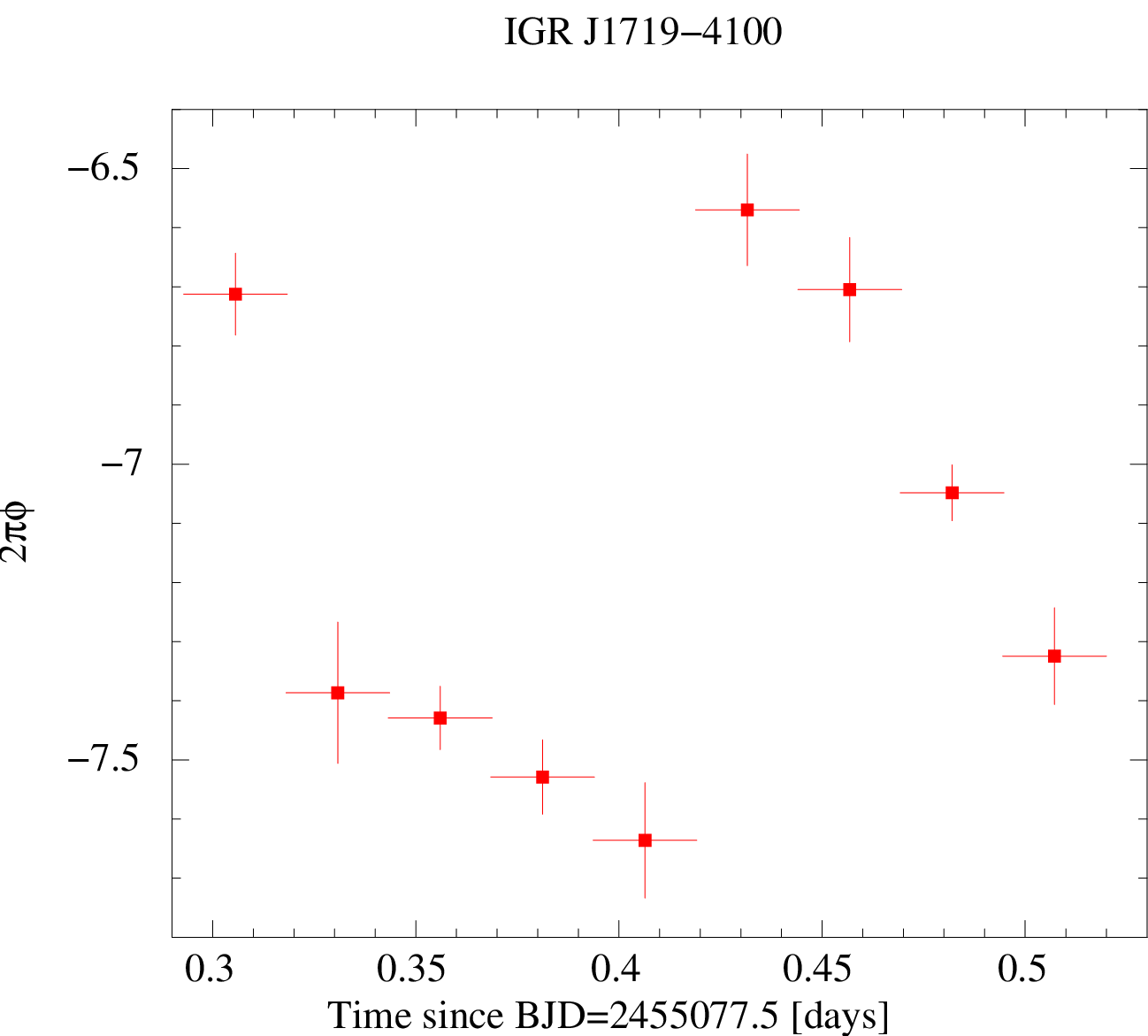}\\
\includegraphics[width=3.0in, height=3.0in, angle=-90]{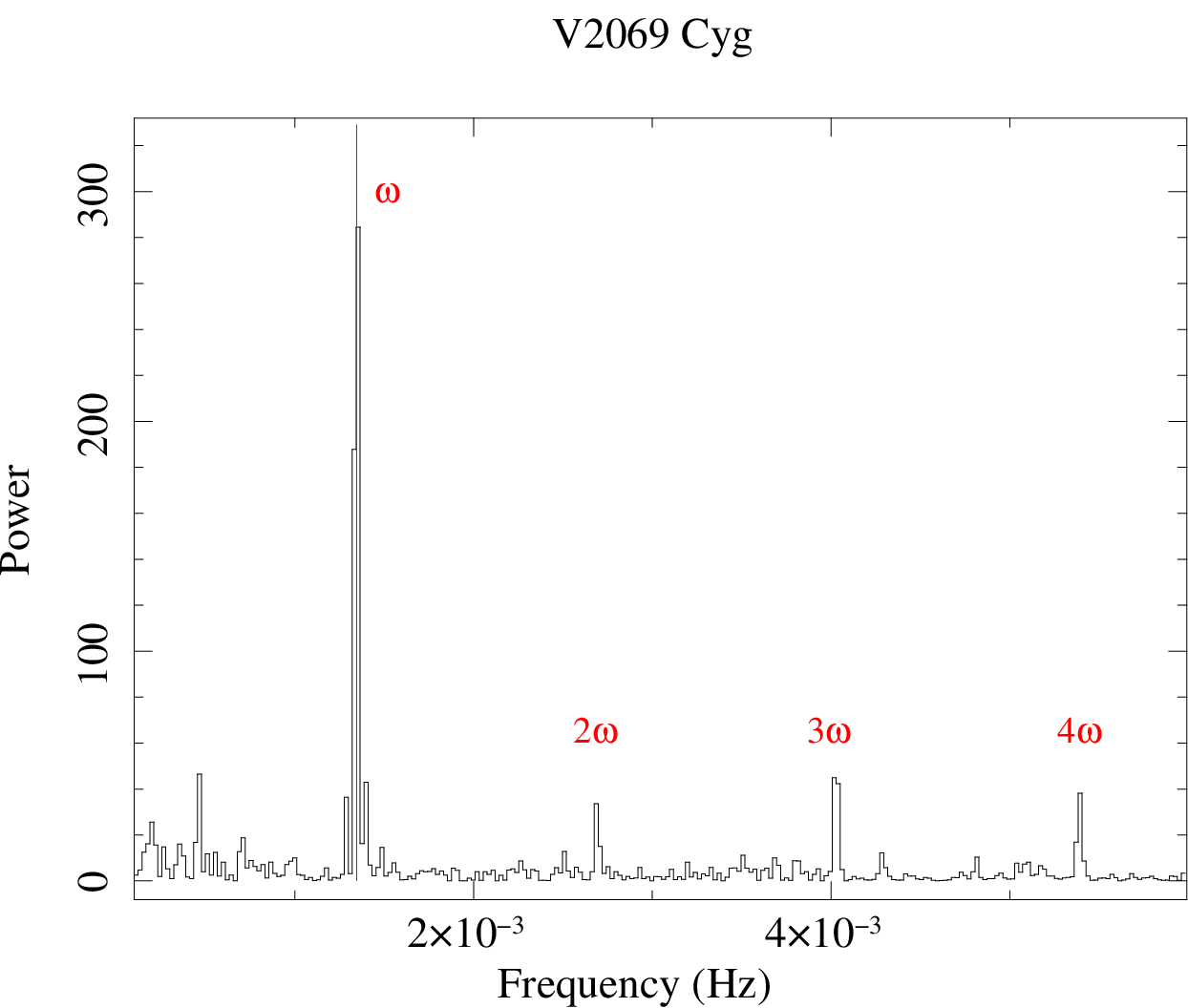} & \includegraphics[width=3.0in, height=3.0in, angle=-90]{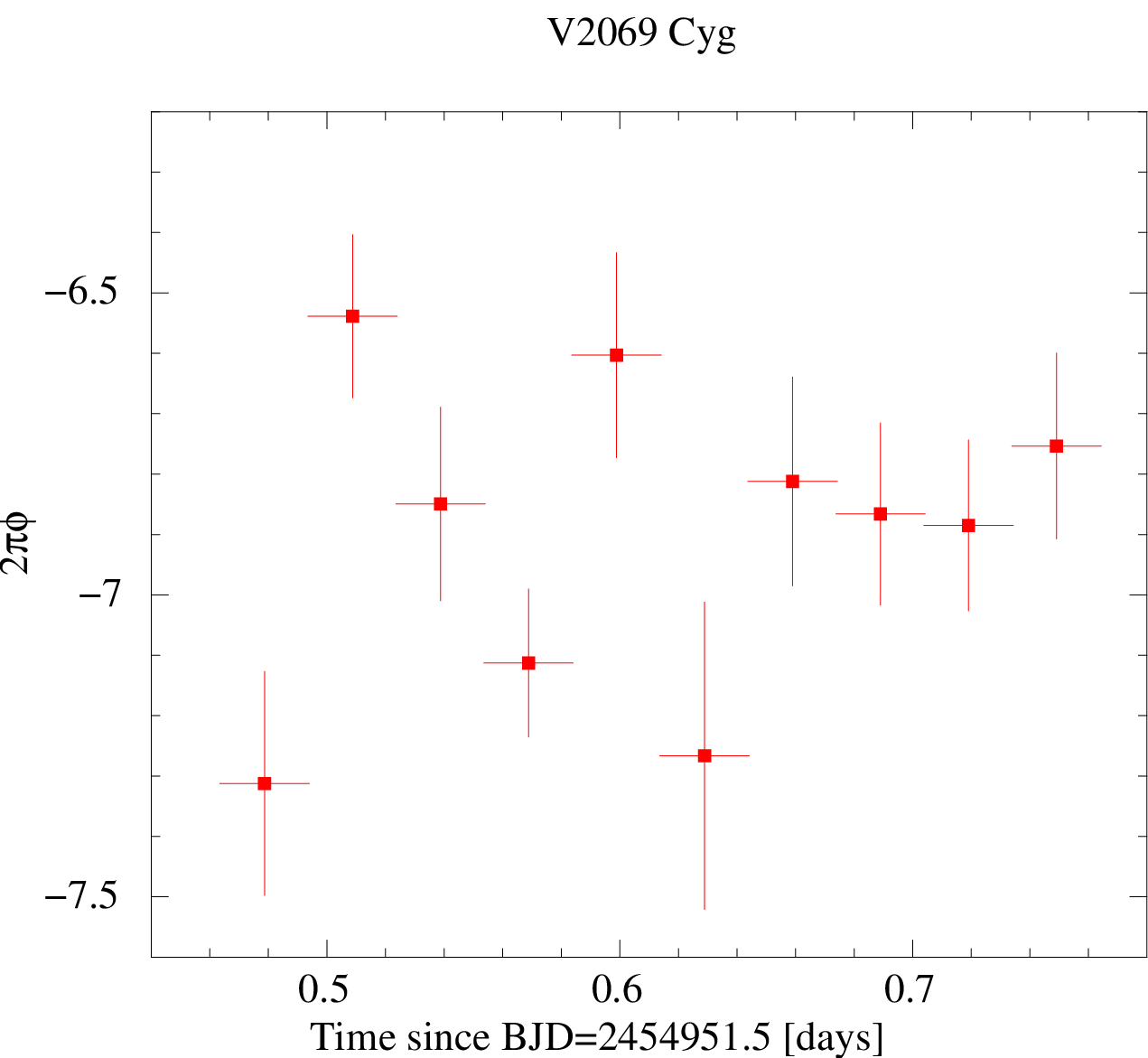}\\ 
\end{tabular}
\caption{As in figure\, \ref{fig:ps_porb1}. P$_{orb}$ with 1$\sigma$ uncertainty is 
also reported for IGR\,J1817.}
\label{fig:ps_porb2}
\end{center}
\end{figure*}

\begin{figure*}
\begin{center}
\begin{tabular}{cc}
\includegraphics[width=3.0in, height=3.0in, angle=-90]{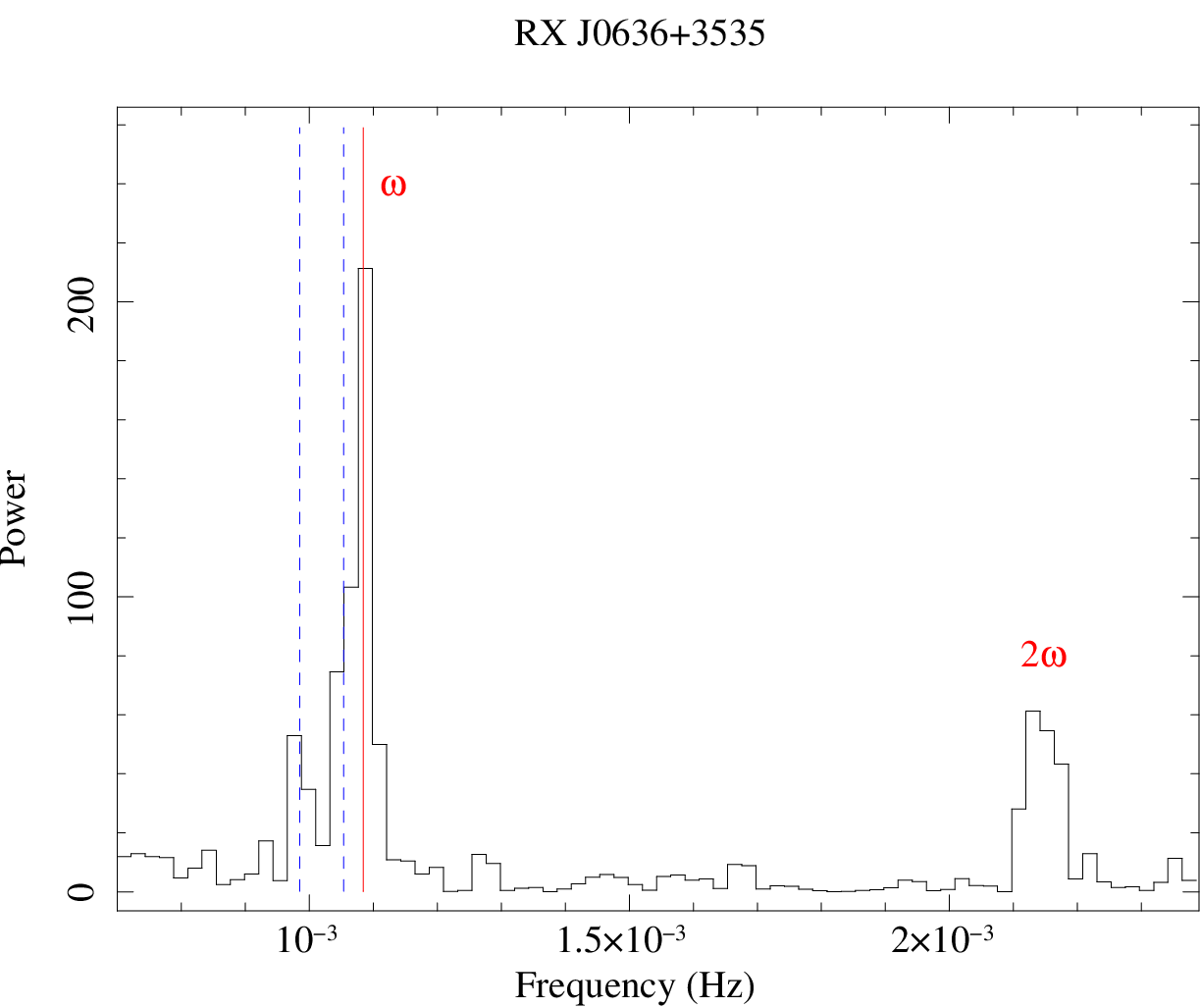} & \includegraphics[width=3.0in, height=3.0in, angle=-90]{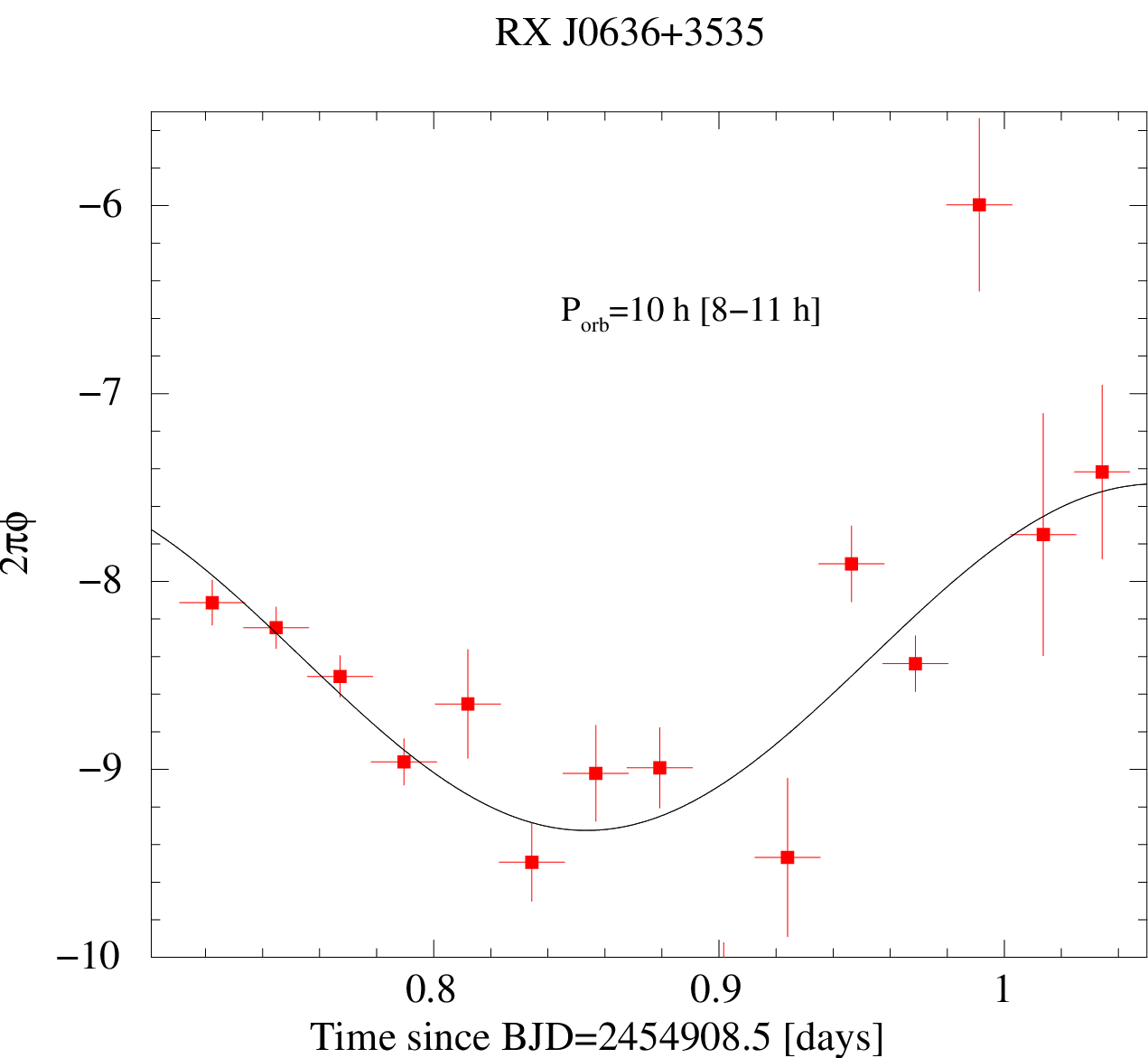}\\ 
\includegraphics[width=3.0in, height=3.0in, angle=-90]{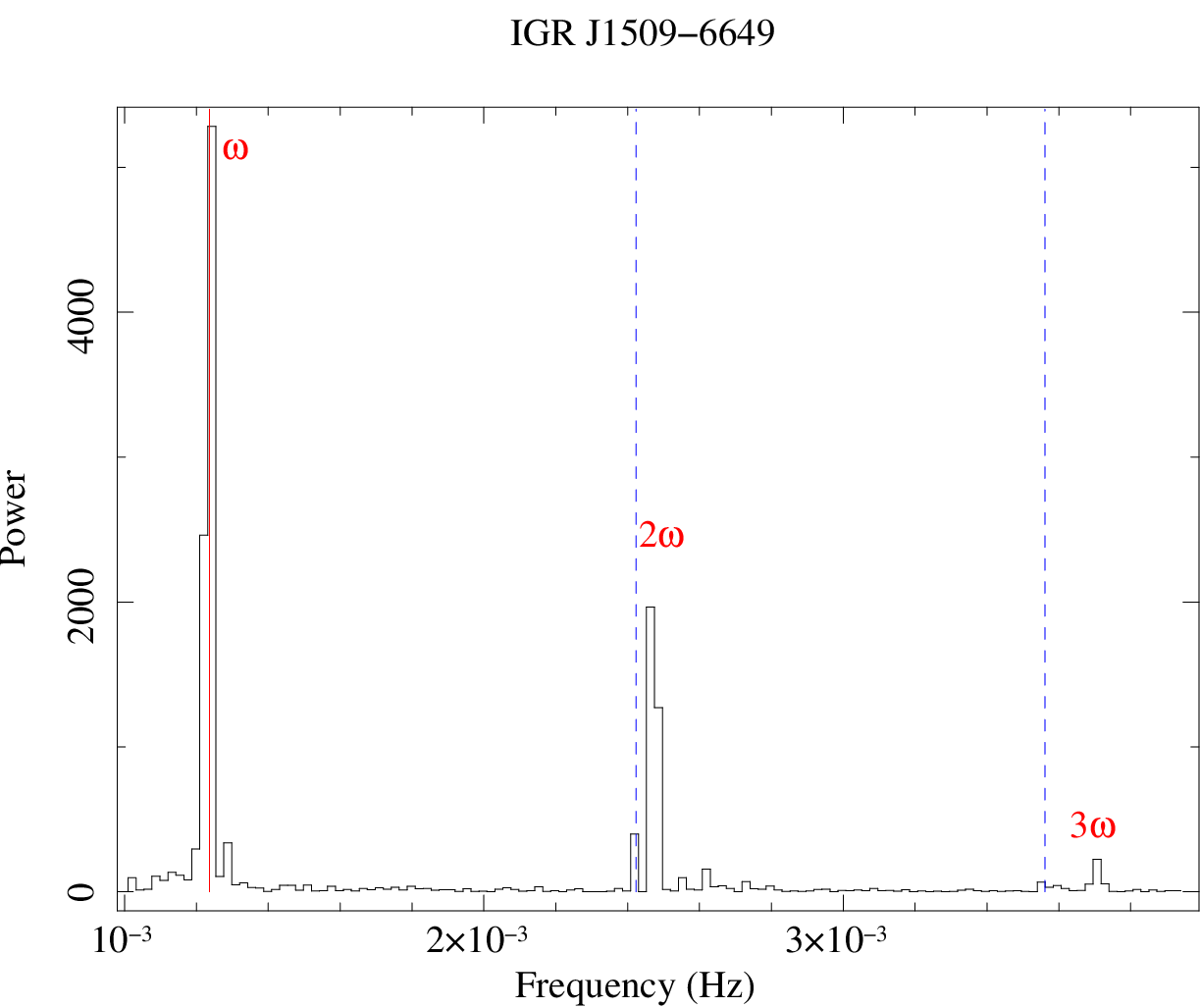} & \includegraphics[width=3.0in, height=3.0in, angle=-90]{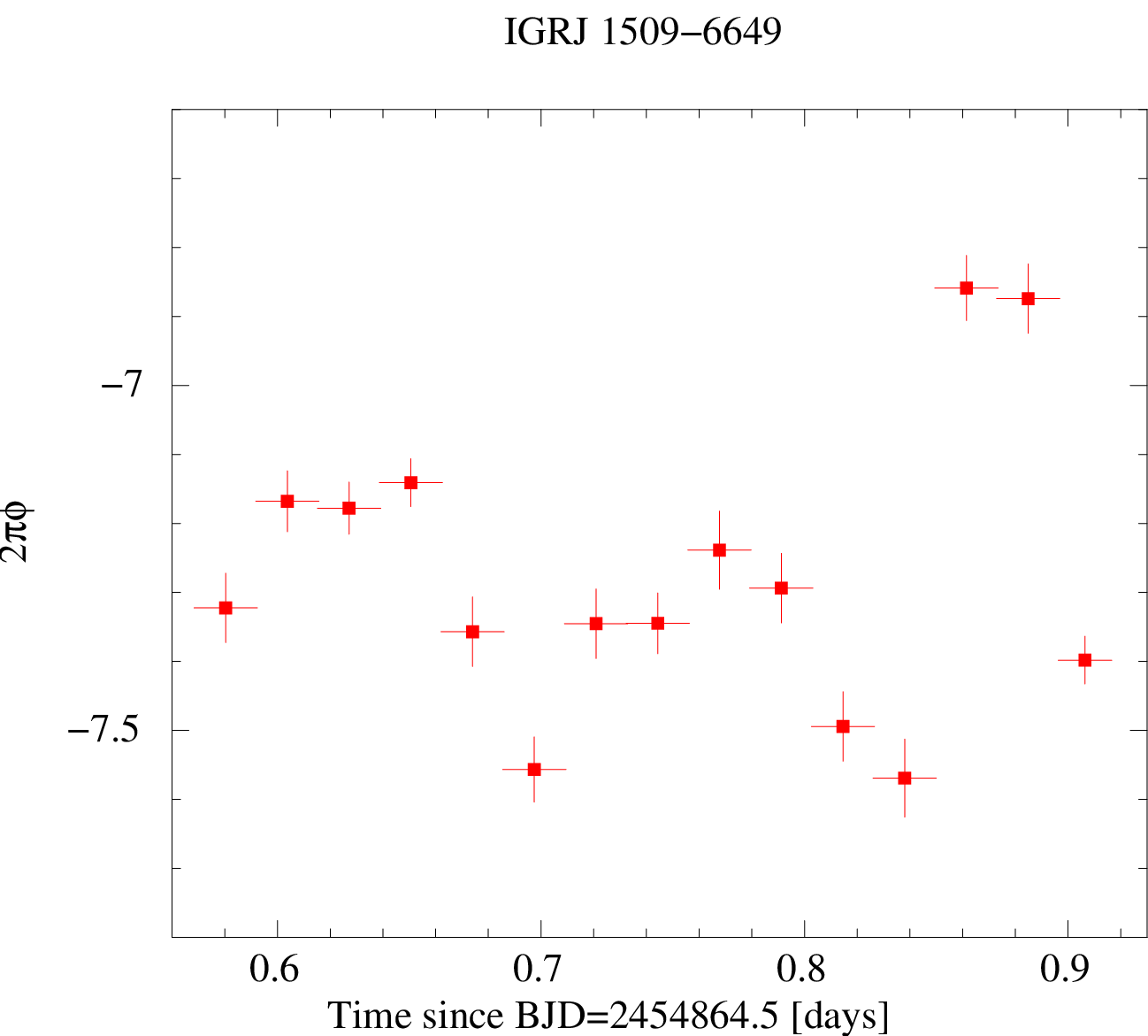} \\
\includegraphics[width=3.0in, height=3.0in, angle=-90]{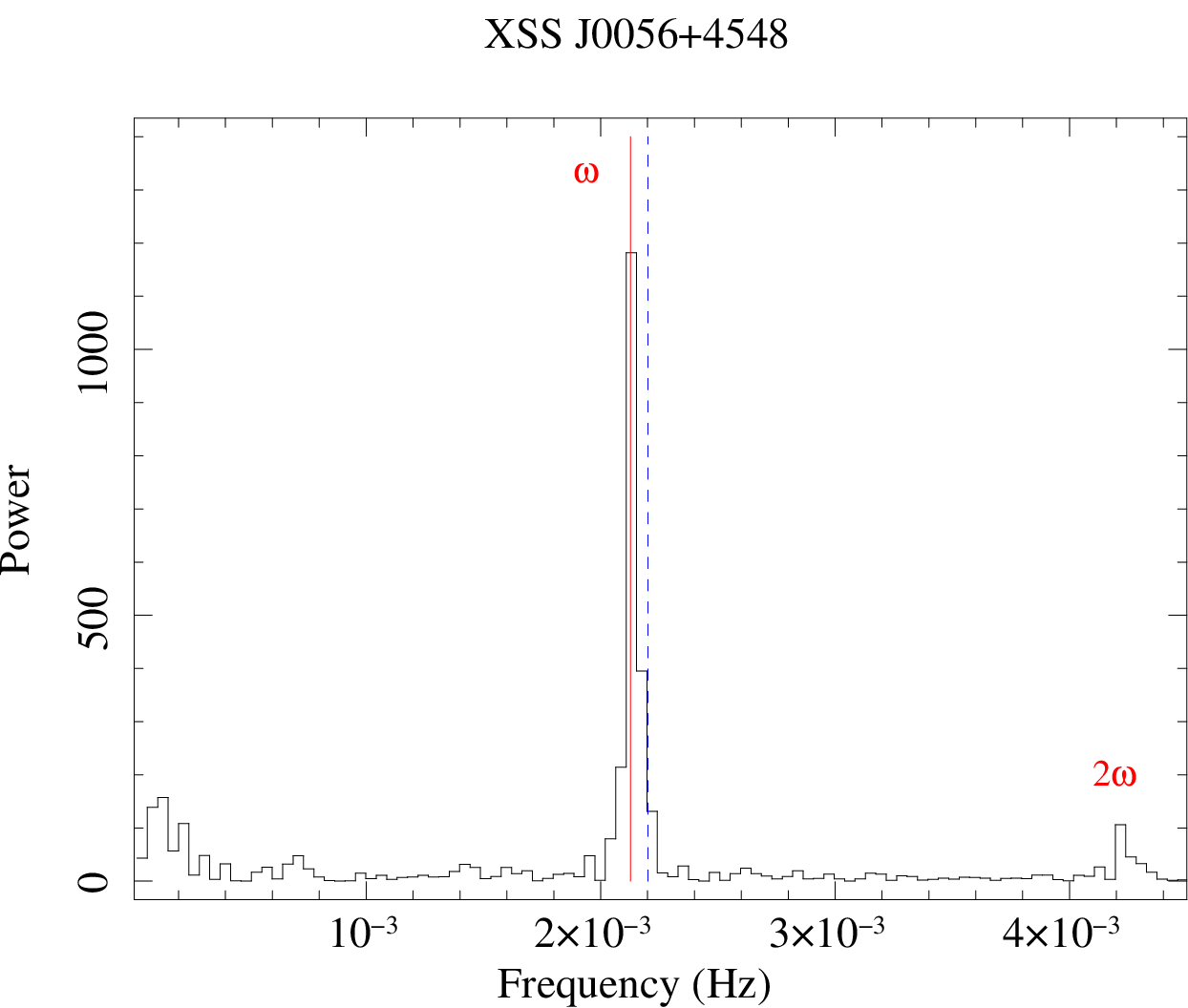} & \includegraphics[width=3.0in, height=3.0in, angle=-90]{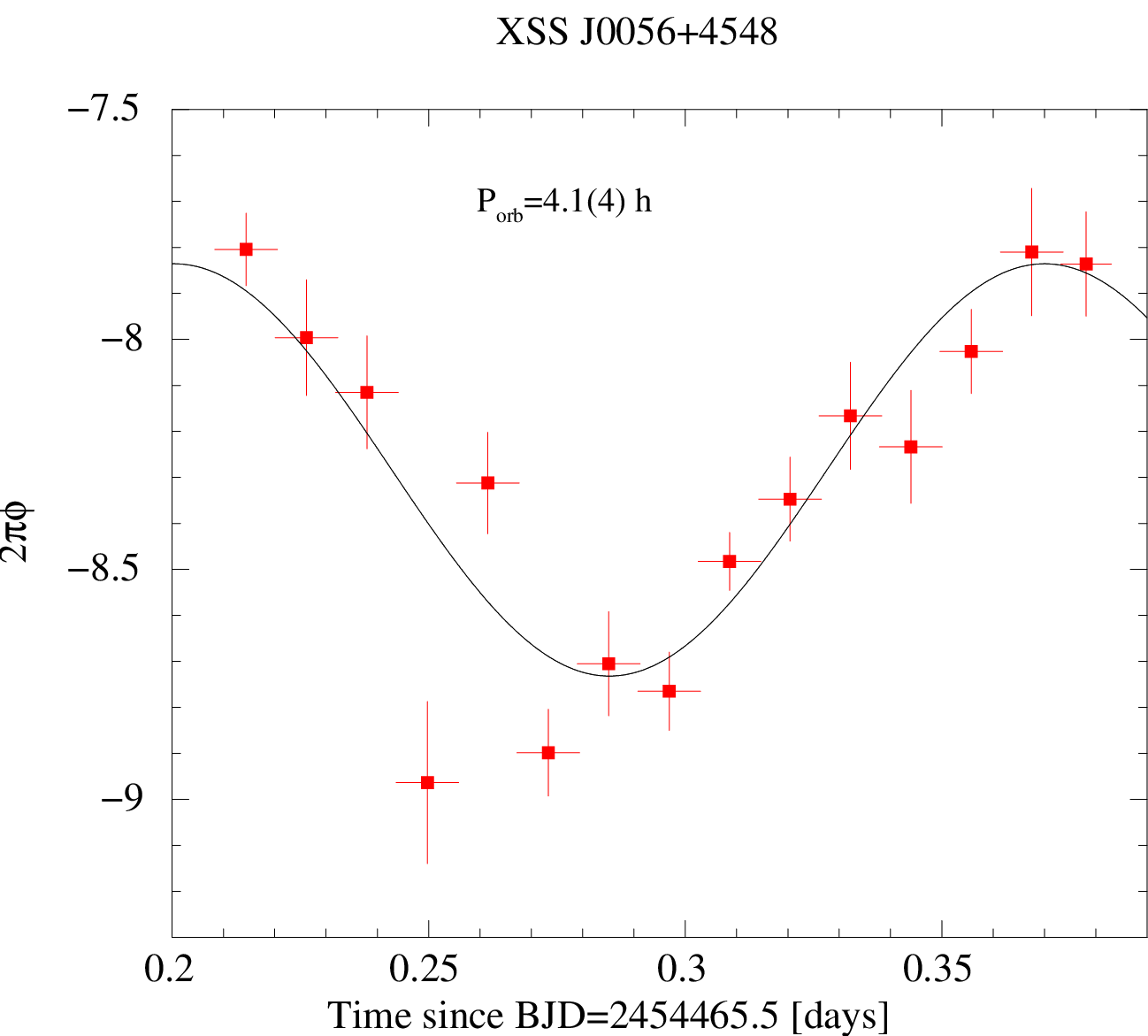}\\
\end{tabular}
\caption{As in figure\, \ref{fig:ps_porb1}.  For IGR\,J1509, the two dotted blue lines refer to $2\omega-\Omega$ and 3($\omega-\Omega$) respectively.
P$_{orb}$ is also reported for RX\,J0636 (with 3$\sigma$ uncertainty) and for XSS\,J0056.}
\label{fig:ps_porb3}
\end{center}
\end{figure*}

We computed the power spectra in the 0.3--15 keV range using the PN data. 
In all of them, strong peaks are found and in many cases also harmonics (see
figure\,\ref{fig:ps_porb1}, \ref{fig:ps_porb2}, and \ref{fig:ps_porb3}). The periods of the 
main peaks were  determined by means of a phase-fitting technique, see 
\cite{dallosso03} for details on the technique. The derived values are reported in table 
\ref{tab:time}. All uncertainties are hereafter at $1\sigma$ 
confidence level (c.l.) if not otherwise specified. For IGR\,J1719 we did not detect 
significant periodic variability in the last third of the observation 
and therefore, the results refer to the first 2/3 only. This source also shows a 
variability on time scale of a few thousands of seconds occurring at about 1/3 of the total 
observation time. This variability is likely responsible for the low frequency peaks seen 
in the power spectrum. We are however unable to characterize it with the present data.

\begin{landscape}
\begin{table}
\caption{Timing properties of the source sample. 
From left to right: P$^X_{\omega}$ (X-ray main (spin) period);
 P$^X_{side}$ (X-ray sidebands); P$^{X,side}_{\Omega}$ (orbital period derived from 
X-ray sidebands); 
P$^{Xpf}_{\Omega}$ (orbital period derived from phase-fitting); 
A$^X_{\omega}$/A$^X_{side}$ (spin to sideband amplitude ratio);
P$^{opt,OM}_{\omega}$ optical spin period from OM; P$^{opt,OM}_{\omega - \Omega}$ (optical 
beat from OM); P$^{opt,spec}_{\Omega}$ 
(optical spectroscopic period with its reference in parenthesis); \textbf{P$^{A}_{\Omega}$ (adopted orbital period in this work.}}
{\small
\begin{center}
\begin{tabular}{cccccccccc}
\hline \hline
   &      &             &         & &               & &  &  \\
Source  & P$^X_{\omega}$ &  P$^X_{side}$  & P$^{X,side}_{\Omega}$ &  
P$^{Xpf}_{\Omega}$ & A$^X_{\omega}$/A$^X_{Side}$ & P$^{opt,OM}_{\omega}$ & 
P$^{opt,OM}_{\omega - \Omega}$ & P$^{opt,spec.}_{\Omega}$&  \textbf{P$^{A}_{\Omega}$}\\
        & s & s  & h & h &   & s   & s & h & \textbf{h}  \\
\hline \hline
\\
IGR\,J0839  & 1480.8$\pm$0.5 & -  & - & - & 
& 1480$\pm$13 & 1560$\pm$7 
& 8$\pm$1.0$^a$& \textbf{8$\pm$1}  \\

IGR\,J1830& 1820$\pm$2 & 2400$\pm$19 ($\omega-2\Omega$) & 4.2$\pm$0.2 & - & 
2.6 &  - & - & - & \textbf{4.2$\pm$0.2}  \\

IGR\,J16500$^b$ & 571.9$\pm$0.5 & -  & - & -  & -  & - & - & 
3.617$\pm$0.003  [1]& \textbf{3.617$\pm$0.003}\\

 IGR\,J1817 & 831.7$\pm$0.7 ($2\omega$) & 810$\pm$1.0 (2($\omega+\Omega$)) 
 & 8.5$\pm$0.2 &  6.6$\pm$0.3 (3.41)$^c$ & 2.5 &  1690$\pm$18 &  -  
& - & \textbf{6.3--8.7}   \\

IGR\,J1719$^d$ & 1062$\pm$2 & 1163$\pm$4 ($\omega-\Omega$) & 
3.2$\pm$0.8 & 
- &  1.4 & - & 
1102$\pm$4 & 4.005$\pm$0.006  [1] & \textbf{4.005$\pm$0.006} \\

V2069 Cyg  & 743.1$\pm$0.6 & -  & -   & - & - & 751$\pm$5 & -   & 
7.48039$\pm$0.00005 
[2]& \textbf{7.48039$\pm$0.00005}\\

RX\,J0636$^e$ & 920$\pm$1  & 952$\pm$4 ($\omega-\Omega$) & 7.6$\pm$0.9   & 
$10\pm^{1}_{2}$ (3.00)$^{c,f}$  &  1.4 & 
- & - & $\sim3.4$  [3] &   \textbf{$\ga$5}\\
              &                        &  1016$\pm$5 
($\omega-2\Omega$)  & 
5.4$\pm$0.5   & & 1.9 &  &  & &  \\

IGR\,J1509$^g$   & 808.7$\pm$0.1 &  842.4$\pm$0.6 ($\omega-\Omega$) & 5.66$\pm$0.08 & - & 
1.4 & 814$\pm$1 & - & 5.89$\pm$0.01  [1] & \textbf{5.89$\pm$0.01} \\
               &               &  412.7$\pm$0.5 (2$\omega-\Omega$) & 
5.6$\pm$0.3  &  - & 2.4 &  - & - & & \\

XSS\,J0056$^h$  & 470.1$\pm$0.2  & 454.2$\pm$0.7 ($\omega + \Omega$)  
& 3.7$\pm$0.2 & 4.1$\pm$0.4 (4.21)$^c$ & 2.6  &  474$\pm$2 & - & 2.624$\pm$0.001
 [4]& \textbf{2.624$\pm$0.001} \\
              & & & 2.52$\pm$0.08$^i$  &  &  & &  & & \\
\hline 
\end{tabular}
\label{tab:time}
\end{center}
\noindent}

$^a$ Orbital period from OM photometric periods.\\
$^b$ Ground based photometric optical pulse at 597.92$\pm$0.04\,s [1].\\
$^c$ Detection significance is reported in parenthesis in unity of $\sigma$.\\
$^d$ Ground based photometric optical pulse at 1139.55$\pm$0.04\,s and 
weaker signal at $\sim1055$ s [1]. \\
$^e$ Ground based photometric optical pulse at 1008.341$\pm$0.002\,s and 
weaker signal at 
930.583$\pm$0.004\,s [3]. \\
$^f$ Since the total exposure is $\sim$8 h, we here give the uncertainty at $3\sigma$ c.l. \\ 
$^g$ Beat period from 3($\omega - \Omega$).
Amplitude ratio refers to 3$\omega$ and 3($\omega - \Omega$). 
Ground based photometric optical pulse at 809.42$\pm$0.02\,s [1] \\
$^h$ OM B band pulses derived from third and fifth segment.
Ground based photometric optical pulse at 470.40$\pm$0.03\,s 
[5].\\ 
$^i$ Orbital period from X-ray light curve. \\

\noindent References: [1] Pretorius 2009; [2] Thorstensen et al. 2001; 
[3] G\"ansicke et al. 2005; [4] Bonnet-Bidaud et al. 2009; [5] de Martino et al. 
(priv.comm.)
\end{table}
\end{landscape}

For each source, the light curve in the 0.3--15 keV range was folded at the main 
($\rm P_{\omega}$) period reported in table \ref{tab:time}, and then fitted 
with a series of sinusoids (one for each harmonic). 
The statistical significance of the inclusion of higher harmonics with 
respect to the fundamental one was evaluated by an F-test 
(see table \ref{tab:pf}). The first harmonic was always 
statistically significant (hereafter we consider significant a 
probability greater then $3\sigma$ c.l.), with the
exception of IGR\,J1830 and IGR\,J1719 in which only the fundamental is
significant.  The second harmonic ($3\omega$) was found to be significant 
in IGR\,J0839, 
V2069\,Cyg and IGR\,J1509. V2069\,Cyg is also the only source showing 4$\omega$.

\begin{figure}
\begin{center}
\includegraphics[width=3.0in, height=3.0in, angle=-90]{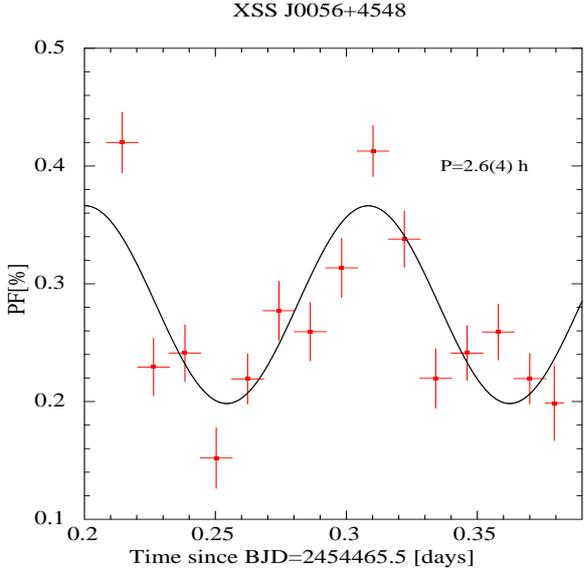}
\caption{PF vs time for XSS\,J0056. The sinusoid period 
with a $3\sigma$ uncertainty is also reported.} 
\label{fig:pfvstime}
\end{center}
\end{figure}

\begin{table*}
\begin{center}
\caption{Pulsed fraction vs energy. Highest detected harmonic and its statistical 
significance is also reported.}
\begin{tabular}{cccccccc}
\hline \hline
\\    
Source        & \multicolumn{5}{c}{Pulsed Fraction}& n$\omega$& sign.\\   
  &0.3--1 keV& 1--3 keV  & 3--5 keV  & 5--15 keV   & 0.3--15 keV& &   \\
        & \%   &  \%  & \%   & \%      &  \%  & &$\sigma$\\
\hline \hline
\\
IGR J0839  & $99\pm2$    & $74\pm1$ & $35\pm2$ & $18\pm2$        &$57\pm1$    & 3      & 3.97  \\
IGR J1830  & $30\pm4$    & $33\pm2$ & $26\pm2$ & $21\pm2$        & $28\pm1$   & 1      &  - \\
IGR J1650  & $18\pm2$    & $12\pm1$ & $5\pm2$  & $6\pm2$         & $9\pm1$    & 2      & 4.29  \\
IGR J1817$^a$  & $58\pm3$    & $46\pm2$ & $34\pm2$ & $27\pm1$        & $30\pm1$   & 2  & 5.94  \\
IGR J1719$^b$  & $14\pm1$    & $13\pm1$ & $6\pm1$  & $\leq3$         &$10\pm1$    & 1  &  - \\
V2069 Cyg  &$14\pm3$     &$25\pm2$  &$16\pm1$  &$9\pm1$          &$18\pm1$    & 4      & 3.5  \\              
RX  J0636  &$18\pm2$     &$21\pm2$  &\multicolumn{2}{c}{$\leq17$}  &$15\pm1$    & 2      & 3.02  \\
IGR J1509  &$57\pm1$     &$51\pm1$  &$27\pm1$  &$9\pm1$          &$42\pm1$    & 3      & 3.52   \\
XSS J0056   &$42\pm2$     &$31\pm1$  &$16\pm2$  &$9\pm1$          &$24\pm1$    & 2      & 4.54  \\
\hline 
\end{tabular}
\label{tab:pf}
\end{center}
$^a$ Pulsed fraction refers to $2\omega$.\\
$^b$ Derived from the first 2/3 of observation.\\
\end{table*}

The power spectra of most sources either show an excess of power
around the fundamental frequency or additional  weaker peaks close-by. In  some cases 
also the harmonics show similar excess or peaks.
Therefore, to estimate their frequency we detrended the light curves from the main pulse by 
subtracting  the composite sinusoidal function corresponding to the 
fundamental frequency
plus their  harmonics when detected. The residual   power spectra of the detrended light 
curves makes clearly visible additional peaks. Consequently, we fitted the 
detrended light curves with one new 
sinusoid (or more sinusoids when harmonics are present). 
We reported the estimated period, when they resulted to be statistically 
significant, in table \ref{tab:time}.
These additional periodicities are identified as the orbital sidebands  
of the main pulse ($\omega$). 

IGR\,J1719, RX\,J0636 show the negative sideband $\omega - \Omega$ 
(beat). In IGR\,J1509 we detected the harmonic of the beat 3$\omega 
-\Omega$, but not the fundamental, and the 2$\omega-\Omega$ 
sideband. In RX\,J0636 we also detected $\omega - 2\Omega$. 
For this source the presence of both negative sidebands is significant at 
4$\sigma$ c.l.. For IGR\,J1830 the $\omega -2\Omega$ is detected at 4.1$\sigma$ c.l., but 
the $\omega-\Omega$ is only at 1.5$\sigma$ c.l..
The positive $\omega + \Omega$ sideband is detected in  
XSS\,J0056 (7.4$\sigma$ c.l.) and 2($\omega + \Omega$) 
in IGR\,J1817  (16$\sigma$ c.l.) since the main X-ray period is the first 
harmonic (see sect. \ref{subsub:optuvar}).
 
We here note that \cite{sazonov08} using a short $Chandra$ observation of IGR\,J0839 
detected a periodicity of $1450\pm40$ s consistent with our more constrained
value.

\noindent \cite{Butters08} failed to detect a periodicity in a RXTE observation  of 
IGR\,J1719. On the other hand, they detected two periodic signals in XSS\,J0056 
again using RXTE. One at 465.68$\pm$0.07 s and a weaker one at 489.0$\pm$0.7 s. The former
is inconsistent with our determination (being out by more than 5 s), while 
the latter is not detected in our data. 
  
\noindent For IGR\,J1817 \cite{Nichelli09} using a Swift/XRT pointing 
report a period of 830.7$\pm$0.07\,s consistent with our results and a period of 
$\sim$1660\,s from a previous short $Chandra$ exposure 
(see sect. \ref{subsub:optuvar} for its interpretation). 

\cite{butters09} by using RXTE data of IGR\,J1509 found a signal at 
809.7$\pm$0.6 s, consistent with our results.

The phase-fitting procedure also allowed us to get an estimate of the 
orbital period. The light curve of each source was subdivided in $n$ intervals (10, 15 or 20), 
where $n$ depends on the source brightness and on the 
intensity of the signal. Then we folded each light curve at the period 
P$^X_{\omega}$ reported in 
table \ref{tab:time} and then we fitted it with a Fourier sine series truncated at 
the highest significant harmonic. 
We consequently studied the variations with time of the phase (and amplitude) 
of the main pulsation, corresponding to the fundamental frequency. 
In most cases the phase variations,  typically within $\Delta \phi \pm$ 0.2, appear to 
be randomly distributed around the average value. 
Only in three sources, IGR\,J1817, RX\,J0636 and 
XSS\,J0056 the lag variations show a clear trend, 
allowing a sinusoidal fit. These are shown in figure \ref{fig:ps_porb1}, 
\ref{fig:ps_porb2}, and \ref{fig:ps_porb3}. The statistical significance of the inclusion of 
a sinusoid respect to a constant function is  evaluated with an F-test, and was found
to be 3.41, 3.00 and 4.21$\sigma$ respectively. The resulting periods are reported in 
column 5 of table \ref{tab:time}.

Most sources show also variations with time of the pulse amplitude
(or PF, see sect. \ref{subsub:pf} for its definition),
but these appear to be randomly distributed around an average value, without 
showing a clear trend. An exception is XSS\,J0056, 
where also the amplitude changes following a sinusoidal modulation.
A sinusoidal fit is statistically significant at 3.52$\sigma$ (see figure 
\ref{fig:pfvstime}), giving a period of $2.6\pm0.4$ h ($3\sigma$ c.l.).
Hence, while in RX\,J0636 and IGR\,J1817 the phase shifts could be due to time 
arrival delays, in XSS\,J0056 the variations seen in both amplitudes and phases 
are likely intrisic. This is also suggested by the presence 
of a peak around $9.3\times10^{-5}$\,Hz in the power 
spectrum and a sinusoidal fit gives a period of 2.52$\pm$0.08 h, in agreement
with the spectroscopic period found by \cite{Bonnet-Bidaud09}.

\begin{figure}
\begin{center}
\includegraphics[width=3.0in, height=3.0in, angle=-90]{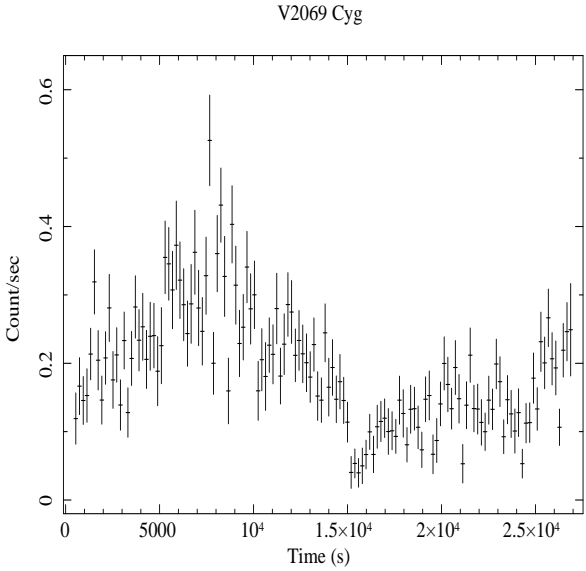}
\caption{0.3--1 keV lightcurve of V2069 Cyg binned at 198 s. Reference time is the 
PN start of exposure, see table \ref{tab:observ}. 
A dip is present at about $1.5\times10^{4}$ s.} 
\label{fig:v2069_dip}
\end{center}
\end{figure}

Interpreting the main pulsation as the WD spin period, 
the sidebands can be used to estimate the binary orbital periods,
reported in  column 4 of table \ref{tab:time}. 
For IGR\,J1719 and IGR\,J1509 the periods are in agreement with
those derived from spectroscopy, reported for comparison in column 9.
The values found for IGR\,J1817 and XSS\,J0056 are consistent within errors with those 
derived from the  phase-fitting method, but for XSS\,J0056  
they  are not compatible with the results from the PF vs time analysis and from the sinusoidal fit.  
For RX\,J0636 we found consistency between the period found from phase-fitting and 
that derived using the beat but not $\omega 
-2\Omega$. However, all derived values are largely off from the tentative period 
of 3.4\,h found from optical spectroscopy \citep{gan05}.
Furthermore, V2069\,Cyg also shows in its power spectrum 
substantial power at low frequencies, but we are unable to constrain a period
from the light curve. However,  a sinusoidal fit with period fixed at 7.5 h, 
derived from spectroscopy by \cite{Thorstensen01}, 
provides a possible match with data and gives a 35$\%$ amplitude modulation. 
This variability is more pronounced at soft energies ($\leq$\,3\,keV),  where we 
also detect a dip, lasting $\sim$1500 s, around the minimum of this long-term X-ray modulation 
(figure \ref{fig:v2069_dip}).

For the systems under study we hereafter adopt, when available,  
the more accurate orbital spectroscopic periods, except for RX\,J0636 for which the orbital period is likely larger, 
\textbf{$\rm P_{\Omega}\ga$5 h}. For those sources without a spectroscopic determination we adopt the orbital 
periods as derived from the X-ray sidebands (for IGR\,J0839 from optical sideband). Exception is IGR\,J1817, for which we adopt the range 
6.3--8.7 h, resulting from sidebands and phase-fitting method.
The adopted values are reported in the last column of table \ref{tab:time}.

\subsubsection{Pulsed fraction}
\label{subsub:pf}

\begin{figure*}
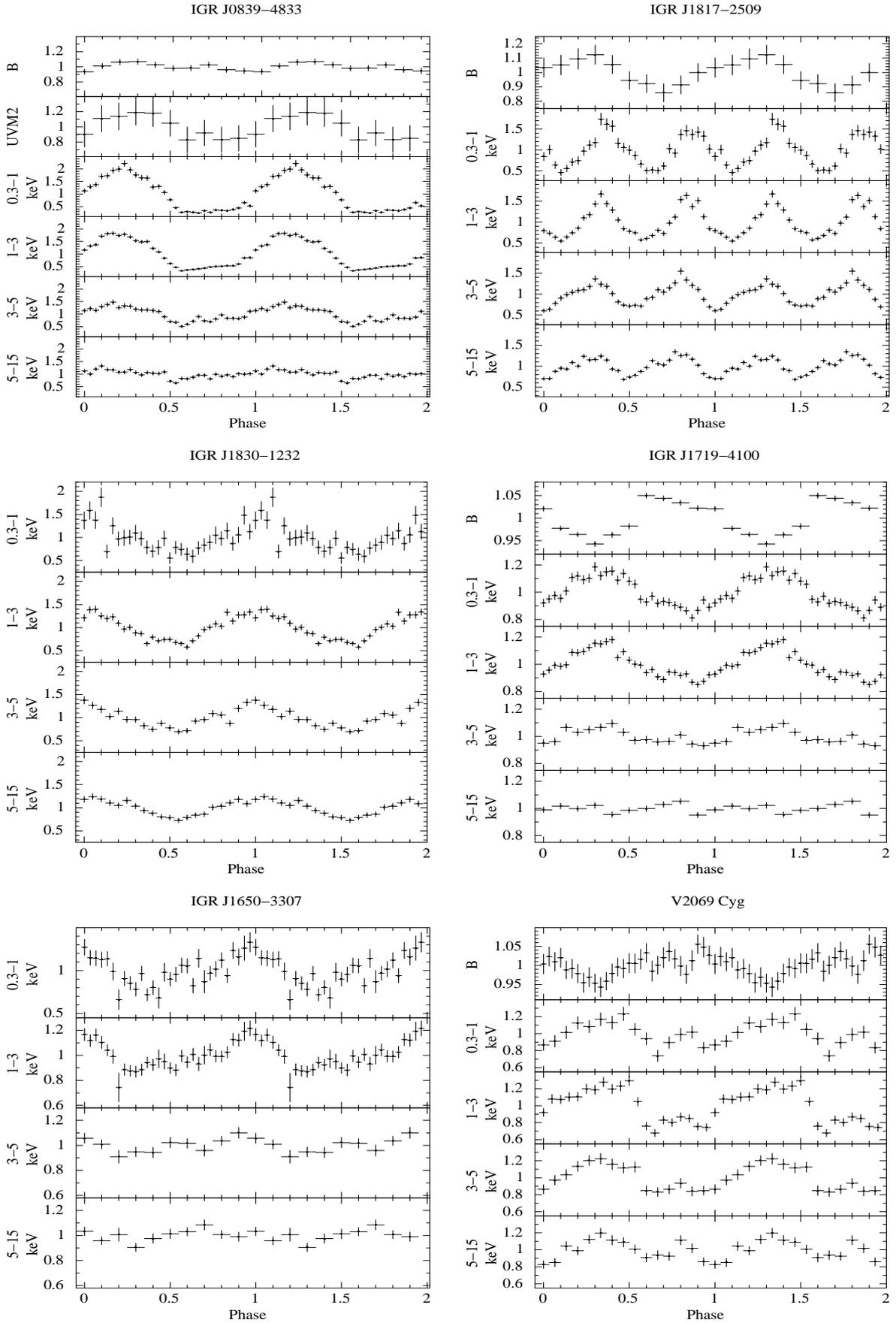

\begin{center}
\begin{tabular}{cc}
\includegraphics[width=3.0in, height=3.0in, angle=-90]{IGRJ0839_final.ps} & \includegraphics[width=3.0in, height=3.0in, angle=-90]{IGRJ1817_final.ps} \\ 
\includegraphics[width=3.0in, height=3.0in, angle=-90]{IGRJ1830_final.ps} & \includegraphics[width=3.0in, height=3.0in, angle=-90]{IGRJ1719_final.ps} \\
\includegraphics[width=3.0in, height=3.0in, angle=-90]{IGRJ1650_final.ps} & \includegraphics[width=3.0in, height=3.0in, angle=-90]{IGRJ2123_final.ps} \\
\end{tabular}
\caption{X-ray pulse profiles at different energy intervals 
for different sources. Energy increases from top to bottom. 
Optical and UV band pulse profiles are 
also reported when available.}
\label{fig:pulse1}
\end{center}
\end{figure*}

\begin{figure}
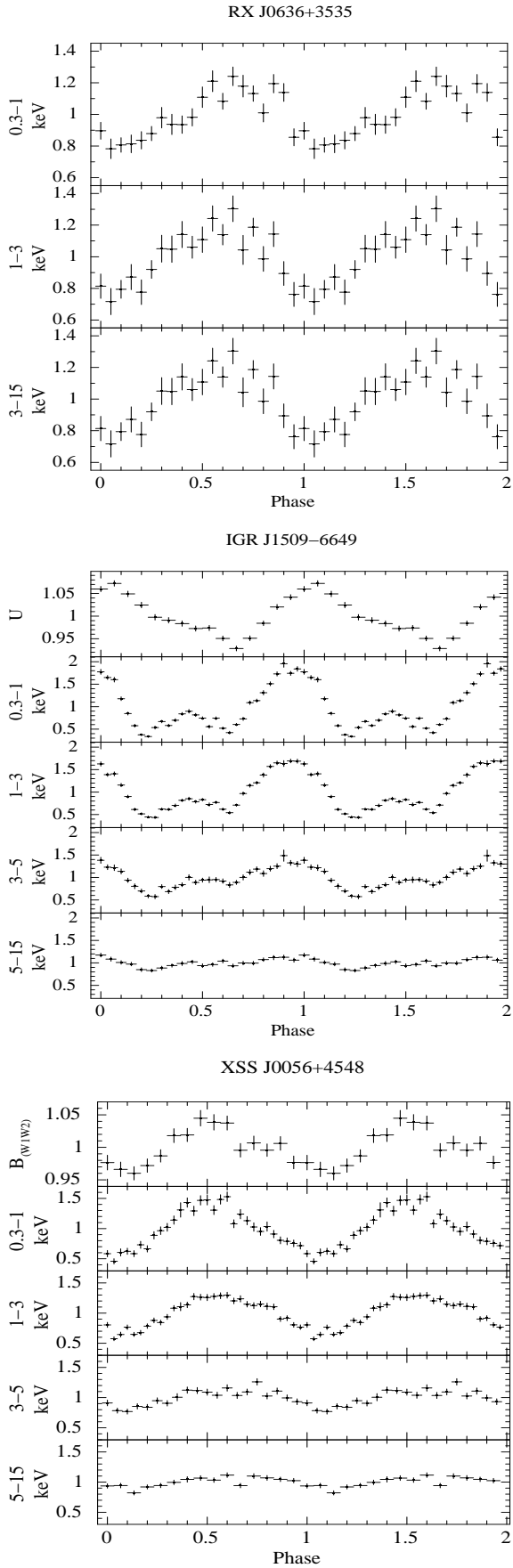

\begin{center}
\begin{tabular}{c}
\includegraphics[width=3.0in, height=3.0in, angle=-90]{RXJ0636_final.ps} \\ 
\includegraphics[width=3.0in, height=3.0in, angle=-90]{IGRJ1509_final.ps} \\
\includegraphics[width=3.0in, height=3.0in, angle=-90]{XSS0056_final.ps} \\
\end{tabular}
\caption{As in figure \ref{fig:pulse1}.}
\label{fig:pulse2}
\end{center}
\end{figure}

We also studied the energy dependence of pulses. We
selected five energy bands, producing four background 
subtracted light curves binned at 66 s (0.3--1, 1--3, 3--5, 5--15 keV), 
plus the 0.3-15 keV light curve with a bin time of 11 s, and folded them at the 
value of $\rm P^X_{\omega}$ reported in table \ref{tab:time}.  
For all sources the X-ray pulse profiles have structured pulse shapes.
 Moreover, for all sources the pulse profiles resulted to be roughly phase aligned at 
all energies, within statistical uncertainties (see figure \ref{fig:pulse1} and \ref{fig:pulse2}).

The folded light curves were then fitted
with a Fourier sine series truncated at the highest significant 
harmonic.
This allowed to study, for each energy interval, the pulsed fraction (PF), here defined as: 
$\rm PF=(A_{max}-A_{min})/(A_{max}+A_{min})$, 
where $\rm A_{max}$ and $\rm A_{min}$ are respectively the maximum and minimum value of the sinusoid used to 
account for the fundamental harmonic. For all sources the PF was found to clearly decrease with energy (see 
table \ref{tab:pf}).
In order to further inspect for possible spectral changes with the spin phase,
we also construct the Hardness Ratios (HRs) using the three energy 
intervals 0.3--1 keV 1--3 and 3--5 keV. 
The HR is here defined as:  $\rm HR=n^{i}_{keV^{a}}/n^{i}_{keV^{b}}$, 
where $\rm n^{i}_{keV}$ is the number of photons detected by the PN camera in the inspected 
energy range (called a and b), and \textit{i} is the phase interval. 
From $\rm HR=n^{i}_{3-5}/n^{i}_{1-3}$ we found 
for all sources a hardening at the pulse minimum, except for IGR\,J1830 
and V2069 Cyg, where HR was found to be constant. 
For V2069 Cyg, due to its peculiar spectrum in 
the soft energy range (E$\lesssim3$ keV; see sect. \ref{subsub:spec}), 
we also inspected the $\rm HR=n^{i}_{1-3}/n^{i}_{0.3-1}$, 
that indicate a softening at spin minimum, 
suggesting the contribution of a soft component.

\subsubsection{The Optical and UV variability}
\label{subsub:optuvar}
Optical and UV light curves were inspected for the presence of periodic 
signals by producing power spectra 
in each band, except for RX\,J0636, IGR\,J1830 and IGR\,J1650 that were
too faint. The UV light curves are generally noisier and for most of them  
we are unable to find significant periodicities. 

The light curves were also fitted with a 
sinusoidal function. The periods were always found 
to be consistent with one of those detected in the X-rays 
(see table\,\ref{tab:time}). Worth noticing are the 
following cases:

IGR\,J0839: The UV light is modulated (PF=35$\%$) at a period ($1480\pm13$ s) 
consistent with that found in the X-rays, but the B band light curve shows a 
period of $1560\pm7$ s (PF=22$\%$). 
The latter is likely the orbital sideband (beat) $\omega$-$\Omega$, 
suggesting an orbital period of $\rm  P_{\Omega}$=$8\pm1$ h. Hence this source is 
spin-dominated in the X-rays and UV, but beat-dominated in the optical.

IGR\,J1817: From the light curve in the B filter we derived a period 
of $1690\pm18$ s, twice the detected
X-ray period. Moreover, no signal was found at the X-ray period.
We conclude that the X-ray period corresponds to the first harmonic, 
while the optical period is the spin.
 
IGR\,J1719: The B band light curve, simultaneous with  
the X-rays, shows a period of $1102\pm4$ s. 
This value  falls in between the two periods detected in the X-rays 
($1062\pm2$ s and $1163\pm4$ s), suggesting that both periodicities 
are also present at optical wavelengths. This is also confirmed by 
ground-based optical photometry \citep{Pretorius09} that showed a weaker 
$\sim1055$ s and a stronger $\sim1139$ s peak in the power spectra. 
Therefore, also this system is spin-dominated in the X-ray, while 
beat-dominated at optical wavelengths (see also sect. \ref{sub:origpuls}).

V2069 Cyg: we derived a pulse period of $751\pm5$ s, 
consistent, within errors, with that found in the X-rays. 
Interestingly, folding the light curve at the X-ray period, the optical 
pulse is anti-phased with the X-ray one.
Moreover, it shows a broad maximum centered on the X-ray minimum, with a dip 
where an X-ray secondary maximum occurs (see Fig\,\ref{fig:pulse1}). 

\begin{figure}
\begin{center}
\includegraphics[width=3.0in, height=3.0in, angle=0]{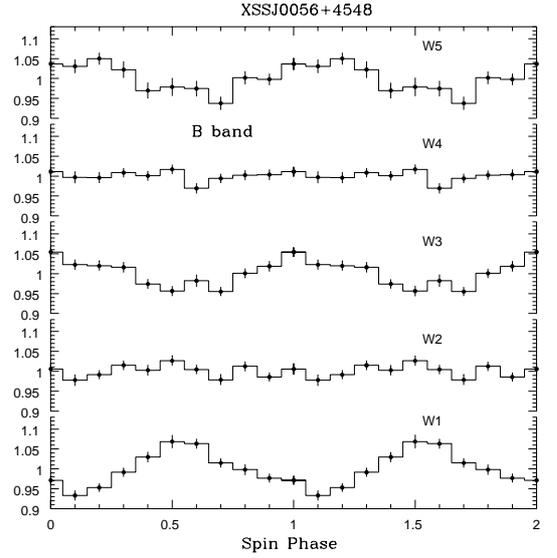} 
\caption{OM light curve of XSS\,J0056 in the B band for the five time intervals of exposure. Time increases from bottom to top.}
\label{fig:figuvopt2}
\end{center}
\end{figure}

XSS\,J0056: it has peculiar UV and optical variabilities.
The power spectra in these bands did not show significant peaks.
However, inspection of the individual five segments ($\sim$1.34ks each)
in the two bands revealed the presence of non-stationary pulses. 
They are found to be phase aligned with the X-ray pulse during the first 2/5 
of the OM B band exposure, for which a sinusoidal fit gives a period of 
474$\pm$4s (consistent with what found in X-rays), 
 but antiphased in the third and fifth segment (see figure \ref{fig:pulse2} 
and \ref{fig:figuvopt2}). The UVM2 filter data revealed
a weak variability ($\sim 9\%$) during the third and possibly
the fourth segment, but we are unable to determine the period. 

\subsection{Spectral analysis}
\label{sub:spec}
\subsubsection{0.3--100 keV spectrum}
\label{subsub:spec}

For all sources, a single-temperature model is inadequate to describe the 
broad band (0.3-100 keV) spectrum.  We found the best match with composite models that 
consist of a combination of a number of optically thin plasma ({\sc mekal}) components 
plus a Gaussian at 6.4\,keV, including a total ({\sc Wabs}) and a partial covering absorber 
({\sc Pcfabs}). The statistical significance of the inclusion of additional {\sc mekal} 
components was always verified with a F-test.
All best-fit spectral parameters are reported in table \ref{tab:averagespec}.

At least two characteristic temperatures were always present, three in the case of 
IGR\,J0839, IGR\,J1830, IGR\,J1650, IGR\,J1817 and  IGR\,J1719.  We broadly divided these optically thin components into 
a cold ($\rm kT_{c} \sim 0.1-0.2$\,keV, exception is IGR\,J0839 with $\rm kT_{c} \sim0.7$ keV), 
a medium ($\rm kT_{m} \sim$4--10 keV) and an hot  ($\rm kT_{h} 
\sim$30--60\,keV) one. The presence of low temperature plasma 
is supported by the RGS spectral analysis of 
IGR\,J1719,  IGR\,J1509, and XSS\,J0056 (see sect. \ref{subsub:rgs}).
 
\begin{figure}
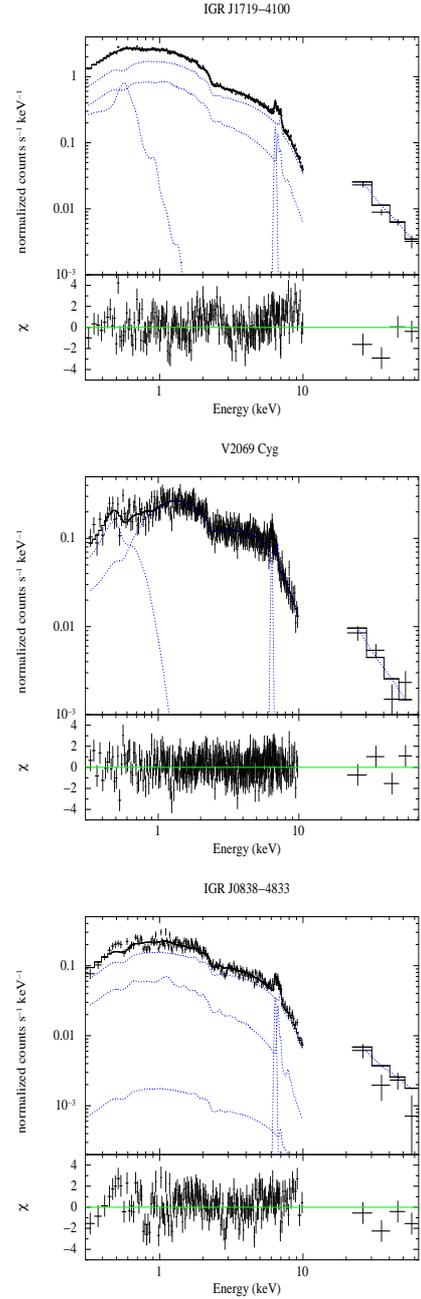

\begin{center}
\begin{tabular}{c}
\includegraphics[width=2.2in, height=2.1in, angle=-90]{IGRJ1719_specfig.ps}\\
\includegraphics[width=2.2in, height=2.1in, angle=-90]{V2069Cyg_specfig.ps}\\
\includegraphics[width=2.2in, height=2.1in, angle=-90]{IGRJ0839_specfig.ps}\\
\end{tabular}
\caption{Broad-band unfolded spectra,  with post fit residuals in 
the lower panels, 
for three representative cases. Spectra have been rebinned for 
plotting purpose. Composite model (solid black line), individual
components (dotted blue lines).
\textit{Up}: IGR J1719.  Wabs*Pc$_{\rm fabs}$*(3mekal+gauss) model used.
\textit{Central}: V2069 Cyg.  Wabs*Pc$_{\rm fabs}$*(bbodyrad+mekal+gauss) model used.
\textit{Bottom}: IGR J0839. Wabs*Pc$_{\rm fabs}$*(3mekal+gauss) model used. An 
absorption feature at $\sim$0.76 keV is present in the residuals (lower panel).}
\label{fig:spec}
\end{center}
\end{figure}

However, in  V2069 Cyg the cold mekal temperature
was unconstrained, while for RX\,J0636, the abundance was unreliably low ($\rm A_Z\sim$0)
A more physically-justified model is obtained by substituting the cold mekal
with a blackbody (BB) component at a temperature of $\rm kT_{BB}\sim$ 
70--80  eV (see also figure \ref{fig:spec}). 

All metal abundances were found to be consistent 
with the solar value except IGR\,J1817, for which   
a highly sub-solar ($0.20\pm0.07$) value was found.

For all sources we found significant presence of an emission 
feature at about 6.4 keV (see figure \ref{fig:spec}), 
with no significant shift in energy. Hence, we fixed the 
centroid energy of the Gaussian component at 6.4 keV. 
The equivalent widths (EW) of this emission feature were found 
$\sim$100-220\,eV,  consistent with  fluorescent Fe  $\rm K_\alpha$ line.  
This suggests the presence of a continuum reflected component, whose
inclusion ({\tt reflect}), however, provides meaningful fit only for 
IGR\,J1650, lowering the temperature of the hot {\sc mekal} (see last 
line of table \ref{tab:averagespec}).
 
All sources  are affected by strong complex absorption.  The hydrogen column density 
($\rm N_H$) of the total absorber was found to be lower or consistent with the 
interstellar value in the direction of the sources. The 
partial absorber, with typical covering fraction $\sim60\%$,  was 
found at much higher column  densities (up to $10^{23}\,\rm 
cm^{-2}$) and hence is instead intrinsic to the sources.  
For IGR\,J1817 and XSS\,J0056, which were found with the highest 
value of intrinsic absorption 
($50\pm10\times10^{22}$ and $300\pm200\times10^{22}$ cm$^{-2}$) 
we used the $Pwab\,\,XSPEC$ model 
 \citep{done_and_magdziarz98}, that assumes a power-law distribution of 
neutral absorbers and is likely more 
appropriate in case of high density intervening absorbing columns. 
This provided a better fit in terms of $\chi^2$.
However, the column density resulted to be poorly 
constrained in XSS\,J0056. 

IGR\,J0839 is the only source which, in addition to the two intervening absorbing columns, 
also displays \textbf{an evident} absorption feature found at 0.76\,keV. 
\textbf{We modeled it using an edge component, which statistical significance, evaluated with 
an F-Test, resulted to be $5.6\sigma$}. This feature is consistent with being 
an OVII K-shell absorption edge with an optical depth $\tau$=0.6. 
It indicates an ionized absorber along the line of sight.

\begin{landscape}
\begin{table}
\caption{Spectral properties for the best fitting model.}
\begin{center}
\begin{tabular}{ccccccccccccccc}
\hline \hline \hline
\\
source   & model     & N$_{H_{\rm W}}$              & N$_{H_{\rm Pc}}$    & cvf      & kT$_{\rm c}$ & kT$_{\rm m}$     & kT$_{\rm h}$       & n$_{\rm c}$ & n$_{\rm m}$     & n$_{\rm h}$      & A$_{\rm Z}$  & EW     &  F$_{0.3-15}$    & F$_{15-100}$  \\
         &           &   $10^{22}$cm$^{-2}$         &   $10^{22}$cm$^{-2}$&  \%      & keV          &  keV             & keV                & $10^{-3}$   & $10^{-3}$       &  $10^{-3}$       &      & keV     &  $10^{-11}$      & $10^{-11}$    \\
\hline 
\\
IGR J0839 &3mek$^a$  & 0.155(6) & 6.8(8)  & 60(2) & 0.69(5) & 6(1)     & 54(9)  & 0.2(1) &1.1(3)       & 3.4(3) & 0.8(1)   & 0.13(2) & 0.75(5) & 1.2(3)   \\
\multicolumn{1}{r}{$\chi^2$/dof 1.13/550}&&&   &&&  &&&&&& \\
IGRJ 1830 &3mek      & 0.82(7)   & 15(2)     & 60(3) & 0.16(1) &6(2)     & 40(20)  & 20(10) &2.1(6) & 4.5(4) & 0.9(2)   & 0.17(2) & 0.94(4) & 0.9(1)          \\   
\multicolumn{1}{r}{$\chi^2$/dof 0.96/359}&&&   &&&  &&&&&& \\
IGR J1650 &3mek      & 0.30(3)   & 11(1)     & 62(1) & 0.18(1) &7(1) & 59(8)  & 0.5(2) &1.4(3) & 7(3) & 1.6(3)   & 0.22(2) & 1.5(1) & 2.7(2)          \\   
\multicolumn{1}{r}{$\chi^2$/dof 1.09/630}&&&   &&&  &&&&&& \\
IGR J1817 &3mek$^b$      & 0.04(2)   & 50(10)$^{c}$ & -0.22(4)$^{d}$ & 0.26(2) &4(1)     & 35(5)  & 9(4) &10(2) & 9.2(6) & 0.20(7) & 0.10(1) & 1.50(1) & 2.4(2)             \\                                         
\multicolumn{1}{r}{$\chi^2$/dof 1.02/731}&&&  &&& &&&&&& \\
IGR J1719 &3mek      & 0.105(3) & 6.3(3)  & 43(1) & 0.159(5) &8(1)     & 30(2)  & 1.5(2) & 7(1)       & 18(1) & 0.62(4) & 0.13(1) & 4.7(2) & 3.8(2)            \\  
\multicolumn{1}{r}{$\chi^2$/dof 1.09/1563}&&&   &&&  &&&&&& \\
V2069 Cyg &bb+mek   & 0.35(4)   & 8(1)  & 64(2) & 0.070(4)$^e$ & - &32(4)    & 7(4)$^f$ & - &6.8(3) & 0.6(2)   & 0.18(2) & 1.18(2) & 1.5(1) \\
\multicolumn{1}{r}{$\chi^2$/dof 1.06/379}&&&   &&&  &&&&&& \\
RX J0636 & bb+mek    & $0.05(1)$   & 11(1)  & 57(3) & 0.079(2)$^e$ & - &36(5)   & 0.8(3)$^f$ & -        & 5.5(3) & 0.6(2)   & 0.14(3) & 1.0(3) & 1.3(3)      \\
\multicolumn{1}{r}{$\chi^2$/dof 1.00/474 }&&&   &&&  &&&&&& \\
IGR J1509& 2mek      & 0.05(1)   & 6.6(6)  & 54(1) & 0.12(1) & -     & 34(3)  & 3(1)  & -        & 10.8(3) & 0.4(1) & 0.17(2) &2.04(2) & 2.3(2)             \\            
\multicolumn{1}{r}{$\chi^2$/dof 1.23/900}&&&   &&&  &&&&&& \\
XSS J0056 &2mek$^b$    & 0.04(1) &300(200)$^{c}$ & -0.60(1)$^{d}$ & 0.16(1) & 10.2(6)     & - &  9.0(4)       & 50(10) &  & 0.42(5)     & 0.13(3) & 3.4(3)  & 1.8(4)  \\
\multicolumn{1}{r}{$\chi^2$/dof 1.03/578}&&&   &&&  &&&&&& \\
\hline \hline
\\
IGR J1650& R*3mek$^g$   & 0.28(2)   & 9(1)     & 60(1) & 0.18(2) &7(1)& 41(4)  & 0.4(2) &1.1(2) & 6(3) & 1.6(fix)   & 0.22(2) & 1.5(1) & 2.8(2)          \\   
\multicolumn{1}{r}{$\chi^2$/dof 1.08/630}&&&   &&&  &&&&&& \\  
\hline \hline \hline                                                                                                         
\end{tabular}  
\label{tab:averagespec}                                                                                                         
\end{center}                                      
$^a$ $E_{edge}=0.75(1)$,$\tau_{edg}=1.3(2)$.\\                                                                   
$^b$ Pwab model used. N$_{H_{min}}$ is fixed at the minimum boundary level of 10$^{15}$ cm$^{-2}$.\\
$^c$ N$_{Hmax}$.\\
$^d$  $\,\,\beta$.\\
$^e$ Blackbodyrad kT.\\
$^f$ Blackbodyrad normalization $\times10^{4}$. It is defined as $\rm n=R^2_{km}/D^2_{10}$, 
where $\rm D_{10}$ is the source distance in unit of 10 kpc.\\
$^g$ Reflection model used. R=0.98$\pm$0.02. Cosine of inclination is fixed at 0.45.\\
\end{table}
\end{landscape}

Since the hot temperature  should be regarded as a lower limit to the maximum
temperature of the PSR, we also used the model of \cite{suleimanov05}, that
takes into account the growth of pressure toward the WD surface and 
hence the change of gravity, to obtain 
a more reliable value for the maximum temperature and consequently an estimate
of the WD mass.  This model is  computed for the continuum only.
We included in the fit the same absorber components as in
the multi-temperature fit, as well as a broad Gaussian that takes
into account the iron line complex. We obtain in all cases
an acceptable $\chi^{2}_{\rm red}$ between 1.1--1.3.
The resulting masses are found in the range 0.74--0.96 M$_{\odot}$ 
(see table \ref{tab:wdmass}). We also note that the temperature of 
the hot component (and hence the
masses) determined with the multi-temperature model are within
$3\sigma$ consistent with those found with the PSR model, except
for XSS\,J0056 whose temperature would give a unrealistic
low WD mass.  This could suggest that the hot component is close
to that at the shock. All other best fit parameters are 
compatible, 
within uncertainty, with that reported in table \ref{tab:averagespec}.
For IGR\,J1650, we also tried to include a reflection component 
without any fit improvement.  This is likely due to the lack of coverage in 
the 10--20\,keV range  and the low statistics above 20\,keV, that prevent 
us from drawing conclusions on the reflected continuum.
 
\begin{table}
\caption{{\it column 2}: WD masses as derived from maximum post-shock 
temperature, using \cite{suleimanov05} model (see text). 
{\it column 3}: Lower limits to the mass accretion rate 
adopting the source distances reported in {\it column 4} and the orbital 
periods in \textbf{\it the last column of table \ref{tab:time}}.}
\begin{center} 
\begin{tabular}{lccc}
\hline \hline 
\noalign{\smallskip}
Source & $\rm M_{\rm WD}$ & $\rm \dot M$ & D$^{*}$ \\
       & $\rm M_{\odot}$ & $\rm M_{\odot}\,yr^{-1}$  & pc\\
\hline 
\noalign{\smallskip}
IGR J0839  & 0.95$\pm$0.08 & 3.2$\times10^{-10}$ & 1200 \\
IGR J1830  & 0.85$\pm$0.06 & 2.4$\times10^{-11}$ & 100  \\
IGR J1650  & 0.92$\pm$0.06 & 4.1$\times10^{-11}$ & 270  \\
IGR J1817  & 0.96$\pm$0.05 & 2.4$\times10^{-10}$ & 610  \\
IGR J1719  & 0.86$\pm$0.06 & 1.9$\times10^{-11}$ & 130  \\
V2069 Cyg & 0.82$\pm$0.08 & 5.3$\times10^{-10}$ & 930  \\
RX J0636   & 0.74$\pm$0.06 & 5.4$\times10^{-10}$ & 1100 \\
IRG J1509  & 0.89$\pm$0.08 & 2.5$\times10^{-10}$ & 600  \\
XSS J0056   & 0.79$\pm$0.07 & 2.5$\times10^{-10}$ & 320  \\
\hline
\end{tabular}
\label{tab:wdmass} 
\end{center} 
$^{*}$ Lower limits to the distances are determined as follows: we used the
lower uncertainty values of the adopted orbital periods  and the 2MASS K band  
(reddening correction not applied) magnitudes of the optical counterparts. For the secondary stars we used 
the K band absolute magnitudes predicted for CV donors as a function of
$\rm P_{orb}$ \citep{knigge06}, assuming that these donors totally
contribute to the K band flux. 
\end{table}

\subsubsection{Pulse Phase Spectroscopy}
\label{subsub:pps}
To understand the role of spectral parameters in 
producing the observed X-ray pulses,
we carried out a phase resolved spectral analysis. 
For all sources we extracted the spectra at the maximum and minimum
of the pulsation (see table \ref{tab:pps} for details on the interval selection). 
We kept fixed the column density of the total absorber, the metal abundance, 
and the parameters of the hot {\sc mekal} component to 
the values found from the  phase averaged spectral fits. We also fixed the 
intermediate mekal temperature, since it was always consistent with 
that derived from 
average fit (except for IGR\,J1650). Furthermore, inspection 
of the  fluorescent Fe line for phase-dependent velocity shifts gave no 
significant change and hence we fixed the Gaussian centroid  at 6.4 keV.
All other parameters were left free to vary (see table \ref{tab:averagespec}).
 
In all cases the spectrum at the minimum is more absorbed than that 
at the maximum (see table \ref{tab:pps}), due to higher values of the hydrogen column density and of 
the covering fraction (however for XSS\,J0056 we fixed the density of the absorber 
to the average spectrum values). 
No variation in the normalization of the cold {\sc mekal} component is found except 
for IGR\,J1650 where it decreases at minimum.
We then conclude that the pulses are mainly due to phase-dependent 
absorption. 

Interesting is the case of V2069 Cyg that instead shows an increase 
in the normalization of the BB component at pulse minimum, in
accordance with the behaviour of the energy dependent pulses (figure \ref{fig:pulse1}). 
This suggests that the secondary maximum at $\phi\sim0.8$, best seen below 1\,keV,
is produced by an increase of the 
normalization of this optically thick component.

We found evidence for phase variation in the EWs of the Fe fluorescent
line only in IGR\,J0839 and  IGR\,J1509. In these sources the EWs 
change between $0.08\pm0.03$ and $0.16\pm0.03$ keV and $0.09\pm0.03$ and 
$0.23\pm0.04$ keV at maximum and minimum, respectively.
               
\begin{table*}
\caption{Spectral parameters of max and min spectra. Other parameters are fixed to their best-fit values of the average spectrum.}
\begin{center}
\begin{tabular}{cccccccccc}
\hline \hline
\\
source                    & N$_{H_{\rm Pc}}$     & cvf      & kT$_{\rm c}$  & n$_{\rm c}$  & n$_{\rm m}$   & n$_{\rm h}$   & EW    &  F$_{0.3-15}$ & $\chi^2$/dof    \\
                          &  10$^{22}$ cm$^{-2}$ &  \%      & keV           & $10^{-3}$    & $10^{-3}$     &  $10^{-3}$    & keV    & $10^{-11}$   &                     \\
\hline 
\\
IGR J0839$^a$   &&&&&&&\\
Max ($\phi=0.0-0.4$)      & 6(1)  & 40(3) &  0.69(fix)   &  0.34(7) & 1.2(3)   & 3.9(3)   &0.08(3) &0.92(3) & 1.20/485 \\                                                                                                          
Min ($\phi=0.5-0.9$)      & 7(1)  &81(1)  &  0.69(fix)   &  $<0.4$  & 0.7(3)   & 3.1(2)   &0.16(3) &0.60(3)  &  \\
\hline
IGR J1830   &&&&&&\\
Max ($\phi=0.9-1.1$)        & 5(2)  & 49(3) &  0.21(1)   &   5(1)  &1.9(3) &  4.7(3)  & 0.22(4) &1.11(6)& 1.03/131\\                                                                                                          
Min ($\phi=0.35-0.65$)      & 14(2)  &65(2)  & 0.16(1)   &   30(10)&1.3(2)  &  3.9(5)  & 0.23(8) &0.76(4) &\\
\hline
IGR J1650   &&&&&&\\
Max  ($\phi=0.8-1.1$)     & 13(2)  & 56(2) & 0.09(1)\,;\,0.7(1)$^b$        & 2.8(4) & 0.10(5) &  8.6(5)    & 0.25(5) &1.58(4) &0.93/353\\
Min  ($\phi=0.2-0.5$)     & 12(1)  & 66(1) & 0.15(4)\,;\,8(2)$^b$           & 0.6(2)& 1.8(5)  &  6.6(7)    & 0.22(3) &1.44(6) &\\
\hline
IGR J1817   &&&&&&\\
Max  ($\phi=0.55-0.75$)   & 50(15)$^c$  & -0.25(8)$^d$     & 0.26(fix)      & 9.6(3)& 15.2(6)  & 10(4)        & 0.15(4) &1.7(1) & 1.06/356\\
Min  ($\phi=0.0-0.2$)     & 80(20)$^c$  & -0.1(1)$^d$      & 0.26(fix)      & 16(7) & 13(9)    &  9(3)        & 0.10(3) &1.4(1) &    \\
\hline
IGR J1719&&&&&\\
Max  ($\phi=0.2-0.5$)     & 6.7(7)    & 39(2)   & 0.18(1)             & 1.2(1)&6(1)  & 19(1) & 0.14(2) & 4.75(3)   & 1.03/1768\\  
Min  ($\phi=0.7-1.0$)     & 7.9(6)    & 47(1)   & 0.15(1)             & 1.7(4)&5(1)  & 20(1) & 0.13(2) & 4.64(3)   &\\  
\hline
V2069 Cyg&&&&&&\\
Max  ($\phi=0.1-0.55$)     & 5.4(6)   & 61(7)   & 0.073(1)$^e$  & 5.8(6)$^f$& - & 7.2(1) & 0.24(8) &1.33(3)    &1.05/275\\                                                                                                          
Min  ($\phi=0.6-0.95$)     & 11(1)    &70(1)    & 0.065(1)$^e$  & 13(1)$^f$ & - & 6.6(1) & 0.27(4) &1.07(4)    &\\
\hline
RX J0636   &&&&&&\\
Max  ($\phi=0.3-0.7$)     & 9(2)  & 49(3) &  0.08(3)$^e$  &  0.6(1)$^f$ &-   &  5.2(4)  & 0.08(3) &1.04(3) &1.08/441\\                                                                                                          
Min  ($\phi=0.8-1.1$)     & 8(2)  &64(3)  & 0.08(1)$^e$ &   0.7(3)$^f$  &-  &  4.7(3)  & 0.07(6) &0.87(5) &\\
\hline
IGR J1509&&&&&\\
Max  ($\phi=0.75-1.1$)    & 5(1)  & 27(2) & 0.11(1)    & 3.2(6) & - &11.6(3)       & 0.09(3) & 2.50(4)        & 1.11/824  \\            
Min  ($\phi=0.2-0.6$)    & 9(1)  & 67(1) & 0.12(1)    & 2.8(6) & - &10.7(3)       & 0.23(4) & 1.82(2)        & \\            
\hline
XSS J0056&&&&&&\\
Max  ($\phi=0.4-0.7$)     & 300(fix)$^c$  &-0.64(1)$^d$ & 0.16(2)           & 8(2)  & 50.0(2) &  - & 0.12(3) &3.2(1)     & 0.96/360 \\
Min  ($\phi=0.0-0.2$)     & 300(fix)$^c$  &-0.52(1)$^d$ & 0.12(4)           & 20(10) & 48.6(2)&  -  & 0.15(3) &2.6(1)     &   \\
\hline \hline 
\end{tabular} 
\label{tab:pps}                                                                                                          
\end{center} 
$^a$ Max: E$_{edge}=0.75\pm0.01$ $\tau=1.4\pm0.2$; Min:  
E$_{edge}=0.76\pm0.01$ $\tau=0.9\pm0.2$.\\
$^b$ Medium mekal temperature.\\
$^c$ N$_{H_{max}}$.\\
$^d$ $\beta$.\\
$^e$ Blackbodyrad temperature.\\
$^f$ Blackbodyrad Normalization $\times10^{4}$. Blackbodyrad normalization is defined as $\rm R^2_{km}/D^2_{10}$, 
where $\rm D_{10}$ is the source distance in unit of 10 kpc.\\
\end{table*}

\subsubsection{RGS spectrum}
\label{subsub:rgs}

The RGS spectra of IGR\,J1719, IGR\,J1509 and XSS\,J0056 (see figure \ref{fig:spec}) were 
fitted in two ways. In one, we used
continuum only models (absorbed BB plus bremsstrahlung)
and fitted the strongest lines using Gaussians, to derive their
centroid, width, and intensity in a manner that does not
strongly depend on the choice of models. In IGR\,J1719,
the lines of OVII, OVIII, NeIX and NeX were all strong enough for
such a treatment; in IGR\,J1509, only the two oxygen lines;
and for XSS\,J0056 only the OVII line (see table \ref{tab:rgs}). 
In addition to the line fluxes, which
are consistent with the prediction of the EPIC model, we found,
for IGR\,J1719 and IGR\,J1509, that the lines are
mildly broadened with Gaussian $\sigma$ of order 1000 km s$^-1$
(or FWHM $\sim$2500 km s$^{-1}$) and that the centroids of the 
He-like triplets (OVII and NeIX for the former and OVII only for the latter) are between
the expected energies of the resonant and intercombination
components, when fitted with a single Gaussian.

We also explored global fits, based on the best-fit
model for the EPIC spectra. For XSS\,J0056, there is
a marginal indication that an addition of a OVII edge
would improve the fit, even though this was not statistically significant
in the EPIC fit.  This should be explored further using
future, higher signal-to-noise spectra.
For IGR\,J1719, although the basic features of the
RGS spectra were reproduced using the EPIC model (including
the normalization), the reduced $\chi^2$ is 1.32 for 363 degrees
of freedom.  We obtained an improved fit ($\chi^2=1.03$, 351 dof)
by including a Gaussian deconvolution using the $XSPEC$ $gsmooth$
model and by letting the mekal density parameter be fitted.
The fit converged to a high density regime (i.e., with no
forbidden components for the He-like triplets) with $\sigma \sim$
1250 km s$^{-1}$, consistent with our findings from the
phenomenological model.  Similarly, a high density, mildly
broadened model was more successful in fitting the RGS
spectra of IGR\,J1509.

\begin{table}
\caption{Spectral parameters with RGS.}
\begin{center}
\begin{tabular}{cccc}
\hline \hline \hline
\multicolumn{4}{c}{IGR\,J1719}\\
Line & E  & $\sigma$  & Flux \\
     &     keV         &   eV      &  photons cm$^{-2}$ s$^{-1}$    \\
\hline
\\
OVII  &  $0.5706\pm_{0.0008}^{0.001}$  & $3.8\pm0.7$           & $(3.3\pm_{0.6}^{6})\times10^{-3}$\\
OVIII &  $0.6516\pm_{0.0013}^{0.0018}$ & $4.3\pm_{1.4}^{1.6}$  & $(7.7\pm_{1.9}^{3.5})\times10^{-4}$\\
NeIX  &  $0.9209\pm_{0.0071}^{0.0078}$ & $11.3\pm_{9.6}^{8.5}$ & $(3.7\pm_{1.2}^{+1.8})\times10^{-4}$\\
NeX   &  $1.0228\pm_{0.0032}^{0.0024}$ & $<32.2$               & $(1.3\pm_{0.8}^{+1.0})\times10^{-4}$\\
\\
\hline
\\
\multicolumn{4}{l}{$\sigma_{g}\,=\,1250\pm_{395}^{515}$ [km/s]}\\
\\
\hline \hline
\multicolumn{4}{c}{IGR\,J1509}\\
Line & E & $\sigma$  & Flux \\
     &     keV         &   eV      &  photons cm$^{-2}$ s$^{-1}$    \\
\hline
\\
OVII  & $0.5705\pm_{0.0015}^{0.0021}$ & $4.1\pm_{1.1}^{1.3}$ & $(7.3\pm_{2.1}^{2.6})\times10^{-5}$ \\
OVIII & $0.6549\pm_{0.0020}^{0.0019}$ & $2.8\pm_{1.1}^{2.8}$ & $(3.6\pm_{0.9}^{1.5})\times10^{-5}$ \\
\\
\hline
\\
\multicolumn{4}{l}{$\sigma_{g}\,=\,1240\pm_{500}^{940}$ [km/s]}\\
\\
\hline \hline
\multicolumn{4}{c}{XSS0056}\\
Line & E  & $\sigma$  & Flux \\
     &     keV         &   eV      &  photons cm$^{-2}$ s$^{-1}$    \\
\hline
\\                                          
OVII  & $0.5676\pm_{0.0444}^{0.0618}$ &  $<240$ & $(3.7\pm_{3.2}^{10.4})\times10^{-4}$ \\
\\
\hline \hline \hline
\end{tabular} 
\label{tab:rgs}                                                                                                          
\end{center}                                                                                     
\end{table}

\begin{figure}
\begin{center}
\includegraphics[width=3.5in, height=3.5in, angle=0]{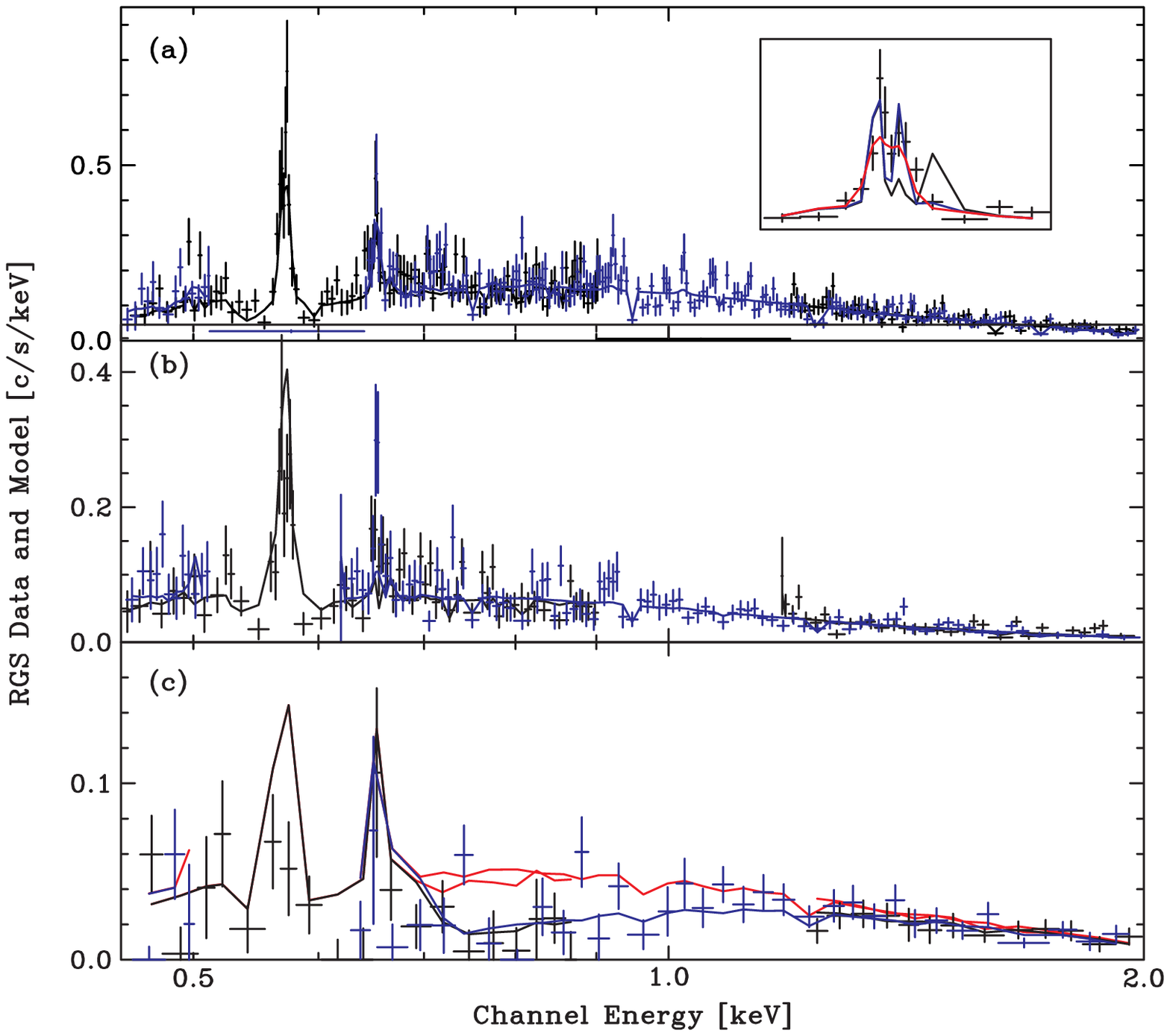} 
\caption{From top to bottom, IGR\,J1719, IGR\,J1509, 
and XSS\,J0056. RGS1 data and model (black), RGS2 data and model  
(blue). The models use the EPIC best-fit values. Those for 
IGR\,J1719 and IGR\,J1509 also include  fitting for
$gsmooth$ $\sigma$ and for $mekal$ density parameters. 
Top panel inset shows the {\sc OVII} line wavelength range 
(20.6--23.1$\AA$) with the EPIC best-fit low density  $mekal$ (black), 
the  high density, narrow $mekal$ (blue) and the high density, $gsmooth$ 
broadened meakl (red).
For XSS\,J0056 we show the EPIC best-fit model with (black) and 
without (red) the edge.}
\label{fig:rgs}
\end{center}
\end{figure}

\section{Discussion}
\label{sec:discuss}
The CVs analyzed here are all found to be strong X-ray
pulsators with periods ranging from $\sim$470\,s to $\sim$1800\,s.
In IGR\,J0839, IGR\,J1817, IGR\,J1719, V2069\,Cyg, IGR\,J1509 and
XSS\,J0056, periodicities are also identified in simultaneous OM optical (B or U 
bands) or UV photometry. X-ray pulses are  unambiguous signature of
magnetically confined accretion and hence of the WD rotational period. Therefore, all 
sources can be classified as CVs of the IP type. 

The \XMM\ data have allowed us: to identify for the first time the WD spin period 
in IGR\,J1830, IGR\,J1650, IGR\,J1817, IGR\,J1719, V2069\,Cyg and RX\,J0636, and to refine  
the spin period of IGR\,J0839 and IGR\,J1509 (but we determined
a slightly different period for XSS\,J0056).  Furthermore, we could infer the orbital periods 
(except for IGR\,J1650). We  provide first estimates for IGR\,J0839, IGR\,J1830, IGR\,J1817 and find consistency with the 
spectroscopic periods for IGR\,J1719, IGR\,J1509, V2069\,Cyg and XSS\,J0056 except 
RX\,J0636. 

We have also presented the first broad band X-ray spectral analysis of these sources
as well as spin phase-resolved spectra, allowing us to 
identify commonalities to IPs, but also peculiarities in some sources.

We discuss here the results in terms of accretion and emission properties.

\subsection{X-ray pulses as tracers of accretion mode}

The presence in the power spectra of a strong and dominant signal at 
the WD spin period, especially in the hard (E$\geq$4 keV) X-rays, is a 
signature that accretion occurs predominantly through a disc 
because the circulating material loses memory of the orbital 
motion \citep{wynnking92}.  For IGR\,J0839, V2069\,Cyg and IGR\,J16500, we have detected only 
the spin frequency in X-rays, and then they likely are pure disc-accretors.  
Noteworthy, the OM optical (but not the UV) data of IGR\,J0839 and the  
ground-based  optical  photometry of IGR\,J16500, IGR\,J1719 
\citep{Pretorius09} and RX\,J0636 \citep{gan05}
reveal instead  stronger signals at the orbital sideband,
$\omega - \Omega$, rather than at the spin period. Since 
the optical light originates from reprocessing of X-rays at fixed regions within the 
binary  \citep{warner}, this makes the X-rays the best tracers of the true 
WD rotational period.

Six out of nine systems, while displaying a dominant spin 
periodicity in the X-ray band, 
 also show signals at sideband frequencies (table\,\ref{tab:time}), with 
spin-to-sideband amplitude ratios 
$\sim$1.4--2.6.  Hence, although the bulk of accretion 
onto the WD takes place through a disc, a fraction of it 
($\sim$35$\%$ up to $\sim$70$\%$) passes over the disc and is directly channeled onto the 
WD magnetic poles. This hybrid mode of accretion, the disc-overflow, is also 
observed in other confirmed IPs \citep{hellier95,norton96,norton97} 
and in more recently identified members \citep{anzolin08}. Worth 
mentioning is the 
case of IGR\,J1719 whose X-ray
light curve reveals spin and sideband pulses 
only during the first 2/3 of observation, while no strong coherent signals are 
found afterward. This system is the one with a low spin to beat 
amplitude ratio of $\sim$1.4 and a spin PF$\sim10\%$. It could be
possible that the two variabilities interfere destructively, 
as also suggested by the large jumps in the phase of the signal with 
time (see figure\,\ref{fig:ps_porb2}) and the low-frequency variations in 
the power spectrum. Hence, the somewhat peculiar behaviour of this IP 
could be due to substantial disc-overflow.

Two systems, IGR\,J1817 and XSS\,J0056, reveal a weak signal at 
a putative positive sideband, $\omega +\Omega$, that would imply a 
retrograde WD rotation. We cannot exclude that the 
true spin period is the latter and hence the stronger signal is the 
beat. In this case the two systems would be beat-dominated at both X-ray 
and optical wavelengths and then disc-less accretor. A similar 
problem raised for BG\,CMi \citep{norton92b,garlick94,demartino95}.  
To solve this issue, both fast time-resolved spectroscopy and polarimetry are 
needed. 

 Furthermore, for IGR\,J1830 and RX\,J0636 the presence of the $\omega 
-2\Omega$ sideband would imply an orbital modulation of the amplitude 
of the beat frequency  if this is an  intrinsic emission. Hence, the beat 
and the spin pulses would be produced by different modes, stream-fed and 
disc-fed accretion respectively, see also \cite{norton97}.

Long-term variation consistent with the orbital period 
was found in XSS\,J0056, and likely in V2069\,Cyg. In the latter, we 
also detected a dip at minimum of the possible orbital 
modulation, more pronounced in the soft bands, indicating a spectral hardening.
Orbital variations are often found in IPs and are likely due to 
absorption effects from fixed regions within the binary frame, such as
the disc edge, in systems viewed at relatively large inclination angles, i$\ga60^o$  
\citep{parker_etal05}.

Using the adopted orbital periods, reported in table 
\ref{tab:time}, the
spin-orbit period ratios, defined as $\rm P_{\omega}/P_{\Omega}$,  are 
lower than 0.1, except for IGR\,J1830 (0.12), and cluster around 
0.03-0.05 (see figure\,\ref{fig:spinorb}). 
  All but one (XSS\,J0056), have $P_{\Omega}$ above the 
orbital gap. A strong asynchronism ($\rm P_{\omega}/P_{\Omega} \leq$ 0.1) 
and long orbital periods are predicted by evolutionary 
models, where mCVs evolve towards shorter orbital periods and, 
depending on the WD magnetic field strength, may or may 
not eventually synchronize \citep{norton04,norton08}. 
In long period highly asynchronous systems, the accretion flow 
is expected to  take part in a disc-like configuration, as indeed inferred 
from X-rays. Such configuration may change while the system evolves towards short orbital
periods, depending on the WD magnetic field strength. To 
further progress in this issue, polarimetric measures of new systems are 
crucial to assess whether these systems will eventually synchronize. 
For IGR\,J1509 a recent polarimetric survey of new IPs has indeed 
revealed that the optical-red light is polarized at a few percent 
($\sim 2\%$), suggesting a strong magnetic field ($\geq$10\,MG) and a 
likely evolution towards synchronism \citep{potter2012}. 

\subsection{Origin of pulses}
\label{sub:origpuls}

In most of the sampled sources the harmonics of the fundamental (up to 
4$\omega$) are detected, indicating departures from strictly sinusoidal pulse shapes. 
Non strictly sinusoidal pulses are seen also in the hard X-rays, where 
absorption is negligible. This suggests that the emitting regions 
depart from a symmetric dipole structure.

V2069 Cyg is the one showing the most structured 
pulse in our sample (see figure\,\ref{fig:pulse1}), with a 
secondary maximum centered on a primary broad minimum seen at all 
energies. This indicates the presence of a secondary emitting pole. 
Furthermore, HRs in the 0.3--1\,keV  and 1--3\,keV bands display 
a softening at spin minimum, indicating an additional contribution at 
this phase. V2069 Cyg also shows a BB component at soft energies (see 
Sect. \ref{subsub:spec}) whose contribution increases at the minimum of pulse (see 
table\,\ref{tab:pps}). Hence, a reprocessed component at the WD 
surface is present and is anti-phased with the optically thin 
emission from the PSR. 

On the other hand, IGR\,J1817 pulsates at a period that is half that 
found at optical wavelengths, indicating two symmetric poles  
contributing almost equally in the X-rays. The optical 
pulse is broad and overlaps the two X-ray maxima 
(figure\,\ref{fig:pulse1}) and hence likely originates 
from wide upper accretion regions above the two poles. 

For all sources the pulses are energy dependent, with 
amplitude decreasing at high energies. The softest band 
(0.3--1 keV) has pulse fractions even up to 
99$\%$, seen in IGR\,J0839, decreasing to a few percent in the harder band 5-15\,keV. 
The pulses harden at spin minimum in most sources as 
confirmed from the spectral analysis (see  Sect. \ref{subsub:pps}).
This is a typical behaviour observed in IP type CVs and consistent with 
the accretion curtain scenario proposed by \cite{rosen88}, where the 
modulation is mainly caused by spin-dependent photoelectric absorption in 
an arc-shaped accretion curtain. The absorption is expected to be larger 
when viewing the curtain along the magnetic field lines (spin minimum). At 
these phases the softest regions are absorbed resulting in a spectral 
hardening.

The curtain is also expected to contribute in the UV and 
optical light and hence the pulses at these wavelengths should be 
in phase with the X-rays. This is the case of IGR\,J0839, and IGR\,J1509, 
though the optical light is dominated by the beat in the first system. 
A peculiar case is XSS\,J0056 that reveals optical pulses 
changing phase with time: they are phase aligned with 
the X-rays only during the first 2/5th of observation. This may suggest a 
non-stationary contribution from two poles in the optical light.
Also IGR\,J1719  and V2069 Cyg display anti-phased X-ray and 
optical pulses that could be due either to a more luminous secondary pole or to the heated WD polar cap. 
For V2069 Cyg the second option could be favored 
because of the detection of the BB component in its spectrum.

\subsection{X-ray emission components and energy budget}
\label{subs:energybud}

The analysis of the average 0.3--100 keV spectra 
gives evidence of multi-temperature emission 
components as well as of complex intrinsic absorption.
A low temperature ($\sim$0.1-0.7\,keV)
plasma traces the regions closer to the WD surface.
The oxygen H- and He-like lines  detected in the RGS spectra of IGR\,J1719 and IGR\,J1509 give further
evidence for the presence of low-temperature plasma. Their intensity ratio  \citep{mauche02} indicates a 
temperature of 
$\sim$0.2\,keV and  $\sim$0.1\,keV, respectively. 
Since no forbidden component is detected in the He-like triplet, 
these lines  form in  high density regime ($\rm n \geq 10^{13}\, cm^{-3}$).
The presence of Ne IX and 
Ne X and their ratio further indicate a 
temperature at $\sim$0.3\,keV in IGR\,J1719. 

The intermediate temperature component is required in the PN spectral
fits to account for the K$_{\alpha}$ lines of Fe XXV (6.7\,keV) and Fe XXVI (6.97\,keV).
These are strong in the spectrum of XSS\,J0056 thus providing a satisfactory fit with only two  
{\sc  mekal} components. 
 
On the other hand, the high temperature component should originate in 
regions close to the shock. The wide range of values 
($\sim$30-60\,keV) found in  our sample would also imply a range of WD 
masses.  The emitting hot regions have typically larger ($\sim$3 times) emission measures 
(EM, normalizations in table  \ref{tab:averagespec}) than  
those of the intermediate component, being that of IGR\,J1650 $\sim$5 
times. As $\rm EM \propto N_H^2\,l^3$, the  larger $\rm EM_{hot}$ 
might indicate a larger emitting volume and hence, on average, a larger 
size. 

Hence, the multi-temperature spectra indicate a temperature 
gradient in the PSR though we are unable to map it with the present data.

Different are the cases of RX\,J0636 and V2069 Cyg that display a 
non-negligible soft BB component with $\rm 
kT_{BB}\sim70-80$ eV. 
These two sources add themselves 
to the small group of soft X-ray IPs \citep{anzolin08} increasing the 
roster to 15 systems. The ratio of bolometric fluxes $\rm 
L_{BB}/L_{th.}$ are $\sim0.85$ for V2069 Cyg and $\sim0.13$ for RX\,J0636, 
respectively. Therefore, this component represents a substantial fraction of the 
energy budget in V2069 Cyg, but not in RX\,J0636. The normalization of this optically thick 
component increases at spin minimum in both sources. It can 
be identified with the X-ray heated WD polar cap, whose projected 
area is larger at this phase. We estimate an average spot area in 
V2069 Cyg, $\rm a_{BB} = 4.5\times 10^{14}\,D^2_{900\,pc}$
and hence a fractional WD area $f\sim 7.6\times 10^{-5}\,D^2_{900\,pc}$\,, for $\rm 
M_{WD}=0.82$ (see table \ref{tab:wdmass} and also its notes for 
the lower limit estimate of distances). For RX\,J0636 we estimate 
 $\rm a_{BB} = 6.3\times 10^{13}\,D^2_{1\,kpc}$ and $f \sim 9\times 
10^{-6}\,D^2_{1\,kpc}$. These values are small, but comparable to what 
found in other soft X-ray IPs with similar BB temperatures. 

An additional component, ubiquitous in magnetic CVs, is the $\rm 
K_{\alpha}$ fluorescent line at 6.4\,keV. It is detected in all sources of our 
sample with large EWs ($\sim$100-200\,eV), arising from reflection of cold material. 
The  lack of velocity shifts and the EW variability at pulse maximum and minimum in 
IGR\,J0839 and 
IGR\,J1509 suggest a likely origin at the WD surface. Such
feature should be  accompanied by a reflection continuum but our data do 
not allow to constrain this component.

A lower limit to the mass accretion rate
can be derived assuming that the accretion 
luminosity is totally emitted in the X-rays: $\rm 
L_{accr} = G\,\dot M_{\odot}\,M_{WD}\,R_{WD}^{-1} \ga 
L_{BB} + L_{th}$, where $\rm L_{th}$ is the bolometric 
luminosity of the X-ray optically thin components and $\rm L_{BB}$ is the 
reprocessed X-ray luminosity, that is only included
for V2069 Cyg and RX\,J0636. The
estimated values are reported in column 3 of 
table \ref{tab:wdmass}. 
The secular mass accretion rates expected 
from angular momentum loss (AML) due to
magnetic braking at the  adopted orbital periods,
would be in the range of $10^{-8}-10^{-9}$ M$_{\odot}$ \citep{McDermott_Taam89}, 
for the sources above the gap. Hence, unless they
are located at a distance 10 times larger than the estimated 
lower limits (very unlikely possibly except for IGR\,J1830 and  
IGR\,J1719), the X-ray emission can not be solely representative of 
accretion luminosity.
A substantial fraction should be reprocessed and emitted in low energy 
bands, as supported  by the detection of periodic signals at the 
spin and/or beat periods at  UV and/or optical wavelengths, 
as firstly raised by \cite{Mukai94}. 
The only exception is XSS\,J0056 that falls in the period gap, for which the 
lower limit to the rate 
is compatible with the values expected either from  
magnetic braking or gravitational radiation \citep{warner}.

The spectral analysis has also shown the presence of dense (column 
densities up to $\sim 10^{23}\,\rm cm^{-2}$) material partially absorbing the 
X-ray emission. The spin phase dependence of absorption  
indicates that the spin pulsations are mainly due to photoelectric 
absorption from pre-shock material in the accretion curtain.
Furthermore, we found indication for the presence of absorbing ionized 
material in one source (IGR\,J0839), whose signature is an OVII absorption edge detected at 
0.76\,keV. IGR\,J0839 adds as the third IP showing this feature 
\citep{Mukai01,demartino08}. A  warm absorber is commonly recognized in 
Low Mass X-ray Binaries  (LMXB) through the presence of absorption edges 
and lines that have been related to ionized atmosphere above the 
accretion disc, giving rise to dips in close to edge-on systems 
\citep{DiazTrigo06}. The spectra at pulse maximum and 
minimum of this source do not provide strong evidence of changing in the edge parameters, 
suggesting that this material could be located within the binary, but not in the 
pre-shock magnetically confined accretion flow. A possible similarity to the LMXB 
dippers should be further investigated with phase--resolved 
spectroscopy along the binary period that we are unable to perform with the present
data.

\subsection{Are hard X-ray IPs different?}

\begin{figure*}
\begin{center}
\includegraphics[width=5.0in, height=5.0in, angle=0]{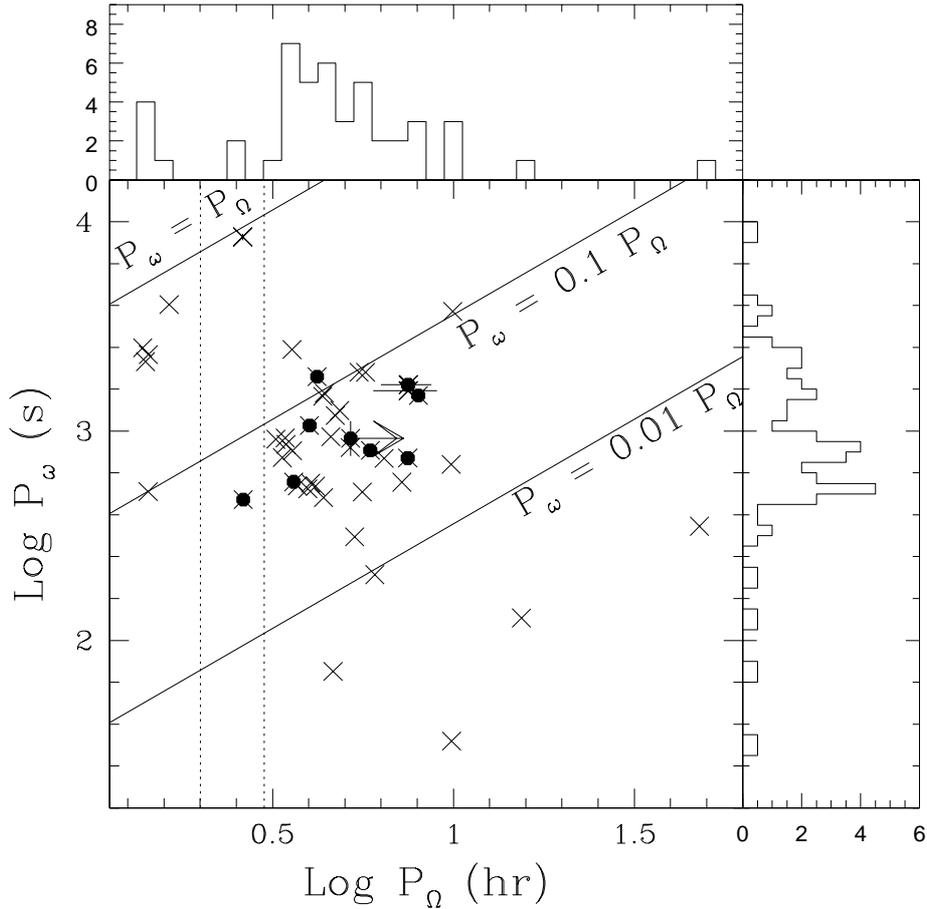} 
\caption{The spin-orbital period plane of confirmed IPs (crosses) including the sample of the nine sources
(filled circles). The ranges of $\rm P_{\Omega}$ for  WX\,Pyx and IGR\,J1817 and the
lower limit for RX\,J0636 are also reported.  $\rm P_{\omega}/\rm P_{\Omega}$=0.01, 0.1 and 1. are
reported as solid lines.  The orbital period gap is shown as vertical dashed lines.  The spin (right panel) 
and the
orbital (upper panel)  period distributions of the whole sample are also shown.  Confirmed IPs are taken from
http://asd.gsfc.nasa.gov/Koji.Mukai/iphome/iphome.html, and we also  include RX J0944.5+0357 \citep{demartino07},
IGR\,J19267+1325 \citep{Evansetal08}, RX\,J052430.2+4244 \citep{Schwarzetal07} and IGR\,J16544-1916
\citep{scaringi11}.}
\label{fig:spinorb}
\end{center}
\end{figure*}

The hard X-ray sources analyzed here are confirmed as IP type CVs.
Their spin periods are well within the range of previously known members 
of this group as depicted in figure\,\ref{fig:spinorb} where our sample 
is shown together with the current confirmed IPs. Orbital periods have been 
estimated combining 
X-ray and optical results.  Most of them are found above the 2--3\,hr 
period gap  except for XSS\,J0056 whose high asynchronism 
($\rm P_{\omega}/P_{\Omega}$=0.05) might suggest 
it will never synchronize. The inferred spin-to-orbit 
period ratios are $\la$ 0.12 consistent with most IPs. Hence, there 
is no striking difference between hard X-ray discovered IPs and other 
known IPs. Nonetheless, the orbital period distribution 
is now being also populated by the new hard systems at $\rm P_{\Omega}\ga$ 5 h.

To inspect whether these selected INTEGRAL and Swift IPs 
possess massive WD we derived their masses using the 
maximum temperature obtained 
from fitting the broad-band spectra with the PSR model 
and taking it as an estimate of the shock temperature (see table 
\ref{tab:wdmass}). These are obtained using 
$\rm T_{shock} = 3/8\,G\,M_{WD}\,\mu_{mH}/k\,R_{WD}$ and adopting the 
WD mass-radius relation from \cite{nauenberg72}. The mean WD mass of 
our sample is 0.86$\pm$0.07 M$_{\odot}$. Consistent results were also found in a 
recent study of 17 IPs made by \cite{Yuasa10}, though they suggest a 
likely bias toward high masses (due to a high energy sampling). 
The average WD mass we found is also fully consistent 
with a recent determination of CV primary masses of 0.8$\pm$0.2 made by 
\cite{zorotovic11}. These authors also find that the WDs in 
CVs are  substantially more massive than those in pre-CVs (0.67$\pm 0.21\rm 
M_{\odot}$)  and than single WDs ($\simeq$0.6\,$\rm M_{\odot}$). While the latter 
cases imply that the WDs increase their mass during the CV 
stage, the fact that IPs  have primaries with masses similar to those of 
the whole CV population, suggests that the mass is not the driving 
parameter. 
Therefore, we conclude that the hard X-ray detected CVs, though  possibly biased toward high 
mass, do not show remarkable difference from the whole CV population.

With the increase in the INTEGRAL and 
Swift exposures with time, most of the undiscovered IPs within $\sim$1.5 kpc will be 
likely detected. This is the case of the faint RX\,J0636, which only 
recently appeared in the new Swift/BAT catalogue 
\citep{cusumano10}. Therefore, the true population
of IP type CVs is expected to be eventually unveiled by sensitive hard X-ray surveys 
of next generation mission. Moreover, as the new discovered IPs 
are being found in the previously poorly populated range of long orbital periods 
(P$\gtrsim$5 h), this will allow us a better understanding of the evolution these systems 
through comparison with recent models, e.g.\cite{Knigge11}. 

\section{Conclusions}
\label{sec:conc}
The main goal of this work is the characterization 
of a sample of 9 new hard X-ray selected CVs, to unambiguously identify 
them as magnetic systems of the IP type. The main results can be summarized as follow:

\begin{itemize}
  \item All sources are strong X-ray pulsators with periods between 470--1820\,s 
and thus are confirmed CVs of the IP type.
 \item  All but two sources are  spin dominated systems 
in the X-rays.  IGR\,J1817 and XSS\,J0056,  
remain ambiguous cases. IGR\,J0839, IGR\,J1719, IGR\,J1650 and RX\,J0636 
are beat dominated at optical wavelengths. 
\item  IGR\,J0839, V2069\,Cyg and IGR\,J16500 are disc-fed accretors, 
while IGR\,J1830, IGR\,J1719, RX\,J0636 and  IGR\,J1509 display a 
disc-overflow accretion 
mode. IGR\,J1817 and XSS\,J0056 could be either stream-fed or disc-fed 
systems.
\item We also estimated the orbital periods (except for IGR\,J1650), all found to be
above the 2--3 h period gap. XSS\,J0056 is the only exception, lying in the gap.
We derived first estimates for IGR\,J0839 ($8\pm1$ h), IGR\,J1830 ($4.2\pm0.2$ h), and IGR\,J1817 (6.3--8.7 h).
\item The amplitude of the X-ray pulse was always found to decrease with energy 
due to complex absorption. 
\item We identified two or three optically thin components with 
characteristic temperatures in the ranges: 0.1--0.7, 4--10 and 
30--60 keV (signature of a temperature gradient in the post-shock region). 
Three are present in six sources.
\item V2069 Cyg and RX\,J0636 are instead found 
to possess a soft X-ray optically thick component at kT$_{BB}\sim$80 eV. 
This increases to 15 the current roster of soft X-ray IPs,
confirming a relatively large ($\sim 30\%$) incidence.
\item A strong K$_{\alpha}$ Fe line at 6.4 keV is 
present in all sources likely originating at the WD surface from the reflection 
of X-rays from cold material.
\item The spectrum of IGR\,J0839 reveals an 
absorption edge at 0.76 keV 
from OVII. This increases to three the IPs were a warm absorber is found. 
\item The WD masses of our sample are found between 0.74 and 
0.96 $\rm M_{\odot}$, with an average mass $0.86\pm0.07\,M_{\odot}$,
suggesting no striking difference with the whole CV population.
\end{itemize}

\begin{acknowledgements}
This publication \textbf{also} makes use of data products from the Two Micron All Sky Survey, which is 
a \textbf{joint} project of the University of Massachusetts and the Infrared Processing and Analysis 
Center (California Institute of Technology), funded by NASA and National Science Foundation.
 FB and DdM acknowledge financial support from ASI under contract 
ASI/INAF I/009/10/0.
\end{acknowledgements}

\bibliographystyle{aa}
\bibliography{biblio}
\end{document}